# Engineering-Economic Evaluation of Diffractive NLOS Backhaul (e3nb): A Techno-economic Model for 3D Wireless Backhaul Assessment

**Edward J. Oughton[12], Erik Boch[3] and Julius Kusuma[3]**
[1]College of Science, George Mason University, Fairfax, VA, USA
[2]Environmental Change Institute, University of Oxford, Oxford, Oxfordshire, UK
[3]Facebook Connectivity, Menlo Park, CA, USA

Corresponding author: Edward J. Oughton (e-mail: eoughton@gmu.com).

This work was supported by an open-science research grant from Facebook Connectivity.

**ABSTRACT** Developing ways to affordably deliver broadband connectivity is one of the major issues of our time. In challenging deployment locations with irregular terrain, traditional Clear-Line-Of-Sight (CLOS) wireless links can be uneconomical to deploy, as the number of required towers make infrastructure investment unviable. With new research focusing on developing wireless diffractive backhaul technologies to provide Non-Line-Of-Sight (NLOS) links, this paper evaluates the engineering-economic implications. A Three-Dimensional (3D) techno-economic assessment framework is developed, utilizing a combination of remote sensing and viewshed geospatial techniques, in order to quantify the impact of different wireless backhaul strategies. This framework is applied to assess both Clear-Line-Of-Sight and diffractive Non-Line-Of-Sight strategies for deployment in Peru, as well as the islands of Kalimantan and Papua, in Indonesia. The results find that a hybrid strategy combining the use of Clear-Line-Of-Sight and diffractive Non-Line-Of-Sight links produces a 9-45 percent cost-efficiency saving, relative to only using traditional Clear-Line-Of-Sight wireless backhaul links.

## I. INTRODUCTION

Task 9.c of the United Nation's Sustainable Development Goals (SDGs) aims to provide universal affordable broadband to all by 2030 [1], [2]. Over many decades, information and communications technologies (ICTs) have been an important enabler of economic development, thus helping to deliver the SDGs [3], potentially lifting millions out of poverty. Hence, solving the digital divide by providing universal and affordable Internet access (SDG 9.c) is critical.

One of the cheapest ways to provide internet access is to use wireless technologies, such as 4G cellular. While the access sites themselves can often be viably built, connecting these assets back into the internet can be a more challenging endeavor for providing coverage, particularly in mountainous areas [4], [5]. The connections between the access sites and the operator's core network are generally called transport links or backhaul links [6].

There are generally three different backhaul technology options [7]. These include fixed fiber optic connections [8],

[9], or wireless methods such as terrestrial microwave [10] and different forms of satellite connectivity, such as low earth orbit (LEO) constellations [11]. Fiber has the highest capacity, while also having the largest capital expenditure (capex) cost [12]. This results from the fact the technology can be slow to plan and deploy, often because local governments require operators to gain public work permits to dig road trenching. Satellite has a lower capex and is much faster to deploy, but data costs are often prohibitive [13], [14]. Microwave offers the best intermediate combination of lower capex (relative to fiber) [10], faster deployment, and lower data costs [15]. In this paper we therefore focus on assessing microwave wireless capex.

On flat plains, wireless links can be used to backhaul traffic over long distances using a single pair of assets (potentially over 100 km, but more commonly below 45 km). Substantial data rates can be provided to users when Clear-Line-of-Sight (CLOS) access is available. However, in situations where CLOS is not possible, Mobile Network Operators (MNOs) have traditionally had to build additional relay sites and 'hops' to help connect remote places back into the nearest fiber Point of Presence (PoP), and the wider Internet [16]–[23]. This additional construction significantly affects the cost of delivery. Subsequently, the use of diffractive Non-Line-Of-Sight (NLOS) backhaul links could help to reduce the costs of deployment [24], potentially enabling many more unconnected users to gain wireless broadband internet connectivity.

A diffractive NLOS backhaul link is defined here as a wireless data connection which utilizes knife-edge diffraction, through which some signal energy is conveyed into the shadow regime of a diffracting feature. In situations where a CLOS link is not possible, the aim is to utilize this approach over shorter distances (e.g. <5 km), thus expanding the feasibility space of microwave backhaul. This is especially useful when trying to traverse between settlements in different valleys divided by mountainous terrain.

Diffractive NLOS links can be implemented using standard microwave backhaul equipment, but the link designer must consider higher signal losses due to diffraction compared to CLOS links [25]. Where appropriate, diffractive NLOS links can be implemented with no change to the network architecture. However, when the appropriate bitrate and bit error rate have been accounted for, the diffractive NLOS link will appear as a regular link that is fully characterized by its bitrate, latency and bit error rate.

Two mobile operators have started to include diffractive NLOS backhaul links into their production networks, providing motivation for this assessment. Internet para Todos ('Internet for All') is a company operating in Latin America, formed via collaboration between the Inter-American Development Bank (IDB) Group, Telefónica, Facebook and the Development Bank of Latin America (CAF) [26]. Mayu Telecomunicaciones is a Peruvian telecoms company also founded to provide Internet to all citizens, regardless of their geography, distance or other market difficulties involved in providing internet services [27].

Planning CLOS links requires large-scale computation of line-of-sight and Fresnel Zone clearance for wireless backhaul design placing larger demands on Three Dimensional (3D) environment models [28]–[32] and pushing the frontiers of geospatial cellular network planning [33]–[40]. CLOS is particularly important for the use of millimeter wave spectrum, over very short distances e.g. <500 m, which will become increasingly common over the next decade due to the large bandwidths being released by governments around the world [41]–[44].

CLOS backhaul is increasingly being proposed for use in small cells networks [45]–[47]. However, NLOS communication applications are also now emerging [48]–[54]. Analysis increasingly shows that high-frequency microwave technology can be used for NLOS wireless backhauling, typically involving frequencies operating below 10 GHz, opening up a range of new wireless applications [55], [56], one of which will be closing the rural digital divide.

Established models exist for *indoor* NLOS (60 GHz) connections but there has been less focus on *outdoor* NLOS applications for mobile cellular communications [57]. Increasingly spatial statistical channel modeling for wireless (4G/5G) backhaul networks is being undertaken for both CLOS and NLOS environments [58], combined with radio propagation measurement, simulation, and analytical results [59]–[61].

Given this background information, the research question outlined for this analysis is as follows:

*What is the cost saving of integrating diffractive NLOS wireless backhaul into least-cost network designs?*

This paper will contribute to the literature in multiple ways. Firstly, a 3D wireless techno-economic assessment method will be presented which advances the field by integrating remote sensing and viewshed techniques. Our aim is to develop a method which can obtain a broad view of the required investment to provide broadband services in a challenging deployment situation, as a precursor to doing detailed modeling on prioritized regions. Secondly, the method will be used to answer the research question and provide, what is to our knowledge, the first openly available techno-economic assessment of diffractive NLOS wireless backhaul strategies.

In the next section a review of network planning approaches is undertaken before a discussion of backhaul technologies is presented in Section III. The method is then articulated in Section IV, before Section V details how this method will be applied to a set of countries to answer the research question. The results are presented in Section VI, with the findings of the analysis discussed, and conclusions given, in Section VII.

TABLE I
COMPARISON OF WIRELESS NETWORK EVALUATION TOOLS

| Eval. level | Name | Purpose | Frequencies (GHz) | Ray tracing? | Terrain and clutter? | Cloud? | Techno-economics? |
|---|---|---|---|---|---|---|---|
| Link | CloudRF [62] | Empirically based propagation modeling | 0.02-20 | No | Yes | Yes | No |
| Link/Cluster | Forsk Atoll[63] | Propagation and channel modeling | 0.02-20 | Yes | Yes | Yes | No |
| Link | Splat! [64] | Empirically based propagation modeling | 0.02-20 | No | Yes | No | No |
| Link | Pathloss [65] | Empirically based propagation modeling | Not stated | No | Yes | No | No |
| Link | Matlab [66] | Propagation and channel modeling | Unlimited | No | Yes | Yes | No |
| Link | EDX [67] | Propagation and channel modeling | 0.04-100 | No | Yes | Yes | No |
| Link | Remcom [68] | Propagation and channel modeling | <100 | Yes | Yes | No | No |
| Link | ANP [69] | Propagation and channel modeling | <100 | No | Yes | No | No |
| Regional | e3nb [70] | Engineering-economic evaluation | 8, 15, 18 | No | No | No | Yes |

## II.  Network planning approaches

When designing a wireless backhaul network there are multiple different types of network planning approaches, with each set of techniques often related to the spatial scale of the assessment area. Fig. 1 illustrates these three areas which include (i) regional level, (ii) cluster level and (iii) link level assessment. Each will now be reviewed in detail.

Technical guidance is provided by the International Telecommunications Union (ITU) for the design of terrestrial wireless backhaul networks [71], including diffractive links [72], although this focuses mainly on the stages carried out in steps two and three (for cluster and link level assessment, respectively). There is generally little guidance provided on the first step focusing on higher level techno-economic business case analysis, with little information shared between operators, as this type of insight is viewed as intrinsic to their own competitive advantage.

Firstly, when developing a greenfield strategy for unconnected locations, high-level evaluation of regional options must be undertaken to help prioritize strategic investments into different areas. Within this process a general understanding of the demand and supply factors which affect the network investment must be quantified. For example, for insight on potential demand, a set of data layers are required to estimate population settlements, along with any available demographic information, to quantify potential revenue. Whereas for insight regarding the supply cost, estimated network designs are required to broadly quantify the engineering cost of delivering potential wireless broadband services.

Secondly, once regions have been selected for investment, cluster level analysis is then required which implements intermediate-fidelity simulations to support network design evaluation [73], [74]. This includes estimating the frequency channels to be deployed along with any bitrate targets aimed at different settlement types [75].

Finally, high-fidelity local simulations are the last step before building the network, with the focus being to optimize the Radio Frequency (RF) engineering parameters to provide the desired quality of service (QoS) [76]–[78], such as the antenna designs, tower heights, potential pathloss and any required margins [79], [80]. There are a variety of software tools available on support these different steps, from regional assessment tools such as the Engineering-Economic Evaluation of Diffractive NLOS Backhaul (e3nb) (presented in this paper), to proprietary local simulation software such as CloudRF, Pathloss etc. as compared in Table 1. Most tools currently available enable CLOS or near-CLOS link planning but do not address the design of NLOS wireless routes.

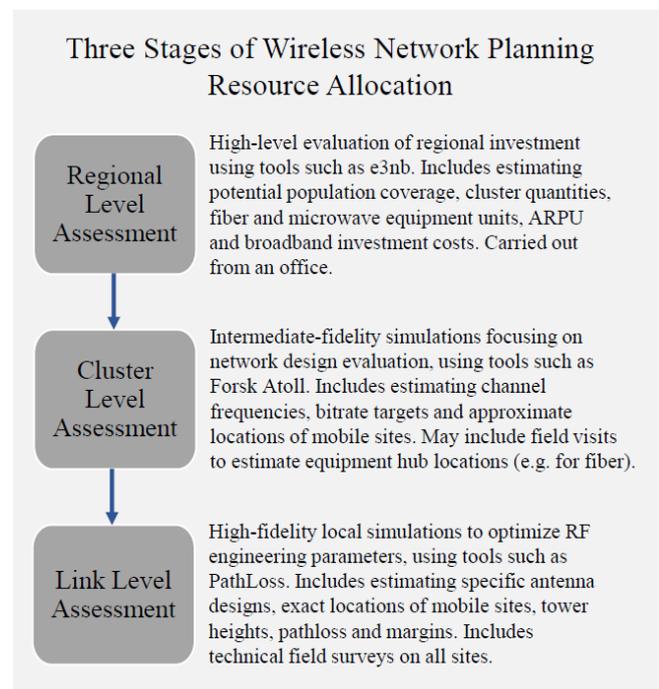

## Three Stages of Wireless Network Planning Resource Allocation

**Regional Level Assessment** — High-level evaluation of regional investment using tools such as e3nb. Includes estimating potential population coverage, cluster quantities, fiber and microwave equipment units, ARPU and broadband investment costs. Carried out from an office.

**Cluster Level Assessment** — Intermediate-fidelity simulations focusing on network design evaluation, using tools such as Forsk Atoll. Includes estimating channel frequencies, bitrate targets and approximate locations of mobile sites. May include field visits to estimate equipment hub locations (e.g. for fiber).

**Link Level Assessment** — High-fidelity local simulations to optimize RF engineering parameters, using tools such as PathLoss. Includes estimating specific antenna designs, exact locations of mobile sites, tower heights, pathloss and margins. Includes technical field surveys on all sites.

**Figure 1** Wireless network planning approaches

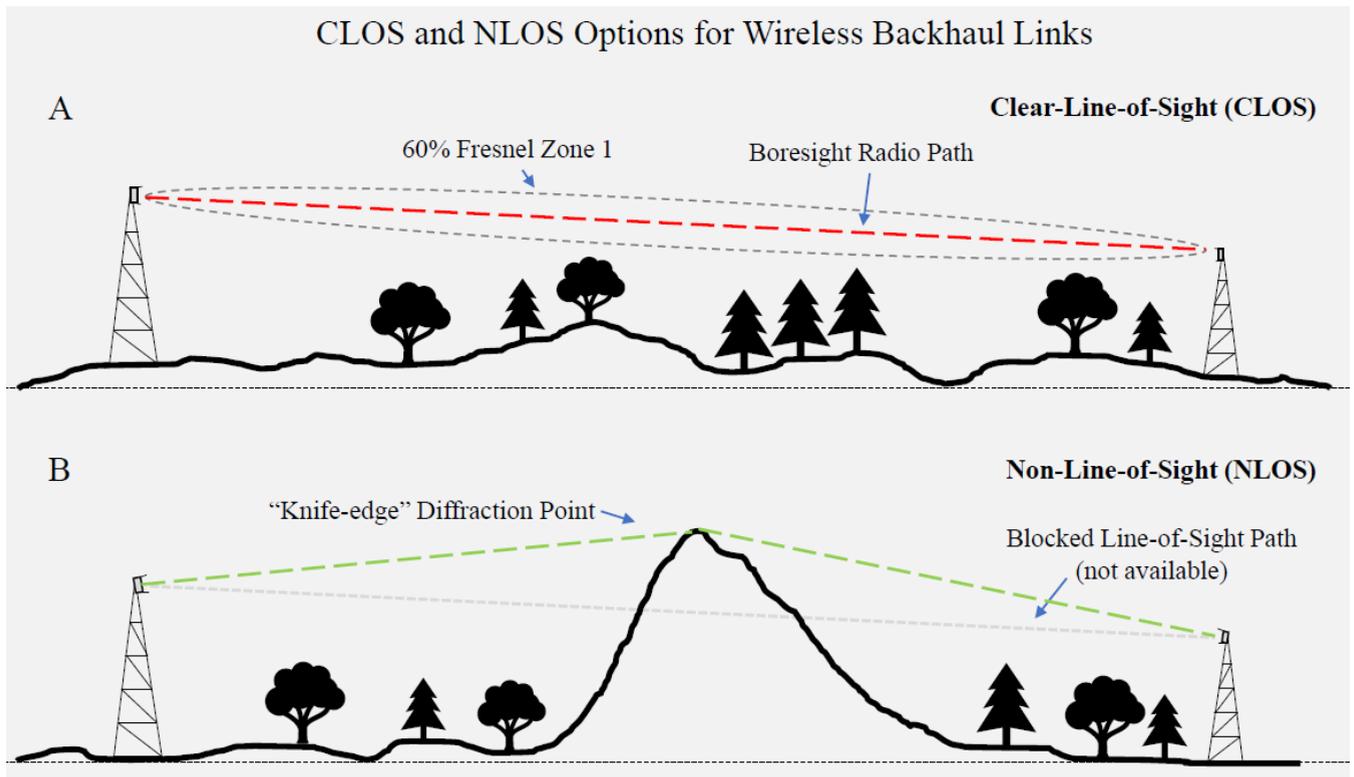

**Figure 2** CLOS and NLOS options for planning wireless backhaul links

## III. BACKHAUL TECHNOLOGIES

There are a variety of backhaul technologies which are used throughout the cellular industry. Recent data from the Global System for Mobile Communications Association (the mobile industry association known as GSMA) estimates that the backhaul technologies in use across different regions can differ quite dramatically, being driven by the income level of the potential users. For example, over 70% of cellular sites in North America have fiber backhaul connectivity, compared to roughly 20% in Latin America [81]. This pattern is reversed when considering wireless backhaul methods, where North America has approximately 25% of towers connected via these means, compared to 80% in Latin America [81]. There are consequential impacts on the quality of service which can be provided to users because of these differences. However, wireless approaches provide a significant reduction in the deployment cost which is useful when the average revenue per user is substantially lower.

Currently, the design and deployment of microwave wireless backhaul overwhelmingly relies on using CLOS to exchange data in a cost-efficient manner over large distances, with well-known design and deployment workflows. For example, see reference [82]. While diffraction is a known phenomenon [72], practical guides caution network designers and engineers that empirically measured diffraction is much worse than analytical prediction. Further, there is only limited material that gives guidance on antenna alignment to optimize link quality in diffracted paths, highlighting the need for further research in this area. As we showed previously via real-world measurements [24], [25], antenna alignment is a very important factor. Fig. 2 (A) provides an illustrative example of this CLOS-focused design-and-deployment workflow. We define the Fresnel Zone clearance as an ellipse shape between two radios which is at least 60% clear of obstruction to ensure satisfactory link performance (as further elaborated in Fig. 4). Common deployment methods involve mounting antenna on erected towers at heights which enable this ~60% Fresnel Zone clearance from the boresight radio path. For those locations which are unable to achieve a direct CLOS, a relay tower (or series of towers) may need to be erected, providing the additional complications of power availability and transportation of necessary materials and equipment.

Recent research has focused on assessing the feasibility of using diffractive NLOS backhaul connections to help drive down the cost of connecting rural and remote locations. Links can 'hop' over knife-edge diffraction points without the need for addition relay towers, as illustrated in Fig. 2 (B).

## (A) Simulation Results for the Received Signal based on the FSPL
### Reported for all frequency and distance categories

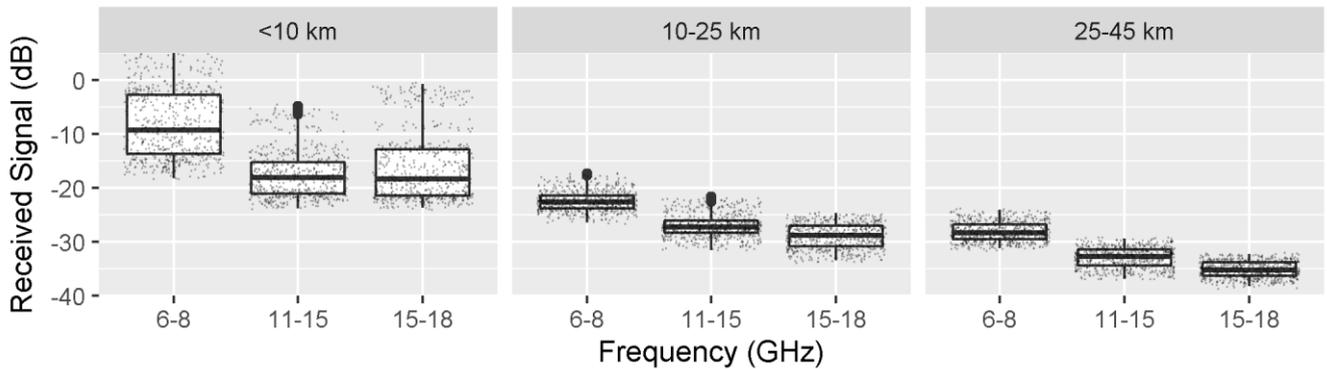

## (B) Simulation Results for Required Fresnel Clearance Values
### Reported for all frequency and distance categories

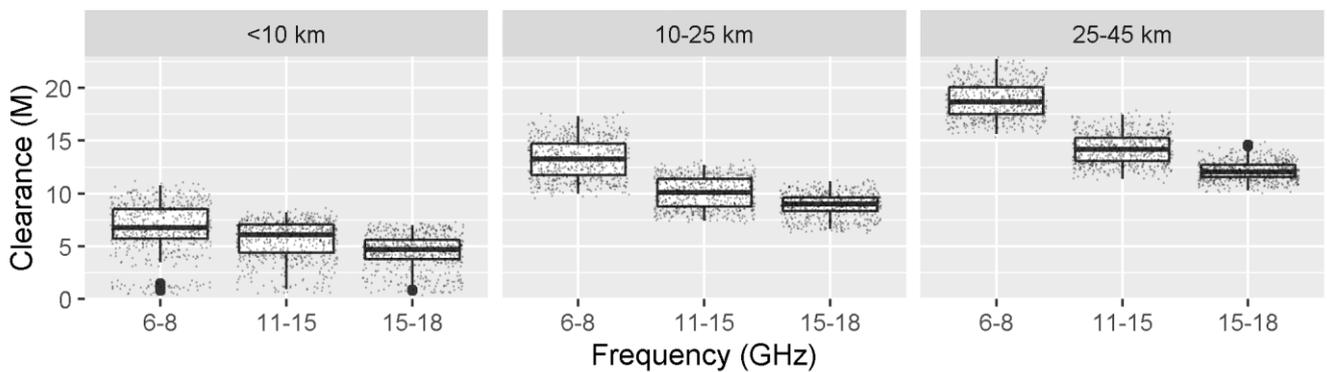

**Figure 3** Received signal and Fresnel clearance by distance and frequency

## IV. METHOD

Two wireless backhaul strategies are to be assessed which include the cost of using either (i) CLOS entirely, or (ii) a mixture of CLOS and NLOS. Network planning methods usually follow a set of deployment rules [83], [84], making it possible to assess the potential effectiveness of different deployment strategies. Firstly, the scientific basis of the assessment approach will be discussed, followed by a description of the model flow for each strategy.

### A. SCIENTIFIC BASIS OF THE TECHNO-ECONOMIC MODEL

Any economic assessment of wireless networks needs to broadly reflect propagation conditions which are dependent on the underlying physics of the electromagnetic spectrum. A surprisingly small number of techno-economic models use engineering techniques to explicitly inform their design, instead usually relying on spreadsheet-based approaches with parameter value assumptions [85], even at the regulatory level [86]. Often this means a lack of initial rigor translates into results uncertainty. To avoid such a situation, we present here a wireless modeling framework which can be used to inform the model input parameters selected.

Firstly, the Equivalent Isotropically Radiated Power ($EIRP_i$) for the $i$th wireless backhaul link can be estimated for any transmitter ($t$) based on (1):

$$EIRP_i = P_t + G_t - L_t \qquad (1)$$

Where the power level ($P_t$), antenna gain ($G_t$) and any antenna losses ($L_t$) can be treated, for example, as 20 Watts, 20 dB, and 4 dB, respectively [87].

Next, the Free Space Path Loss ($FSPL_i$) for the $i$th wireless backhaul link can be estimated for CLOS channels [88], based on the distance ($d$) between the transmitter and receiver (in km) and the carrier frequency ($f$) of the spectrum band (in GHz), as per (2):

$$FSPL_i = 20 \log(d) + 20 \log(f) + 32.44 \qquad (2)$$

Finally, the received power ($RP_i$) for the $i$th wireless backhaul link can be estimated at the receiving antenna, as per (3).

$$RP_i = EIRP_i - FSPL_i + G_r - L_r \qquad (3)$$

Where the receiver ($r$) has an antenna gain ($G_r$) and antenna losses ($L_r$), for example, which can be treated as 20 dB and 4 dB, respectively [87]. Thus, for different backhaul frequency ranges, the received power can be visualized against the potential distance between the antennas, as reported in Fig. 3 (A). In line with wireless communications theory, this demonstrates that the received power is higher for lower carrier frequencies and vice versa. Moreover, the received power values drop rapidly in the first 10 kilometers traveled, and then more modestly between 10-50 kilometers.

Documentation for industry-standard wireless backhaul equipment suggests a 100-200 Mbps link can broadly be achieved with a received signal above -55 dBm and 50 MHz channel bandwidth [89]. Combined with Fig. 3, such insights can be used to guide the development of the techno-economic model inputs.

These parameters generally vary by geography, desired link availability, antenna size limitations, regulatory factors, and therefore are best represented in terms of typical ranges. Moreover, this does not account for environmental clutter (buildings, trees etc.), interference, rain attenuation etc. In practice, all these factors would be considered on a case-by-case basis, at either the cluster or link level planning stages, using the predictive planning tools outlined earlier in this article in Table 1, when such detail is required.

Next, one of the major factors needing to be assessed is the calculation of the maximum required Fresnel clearance [31] above the tree canopy present, for the $i$th wireless backhaul link, as visualized in Fig. 4.

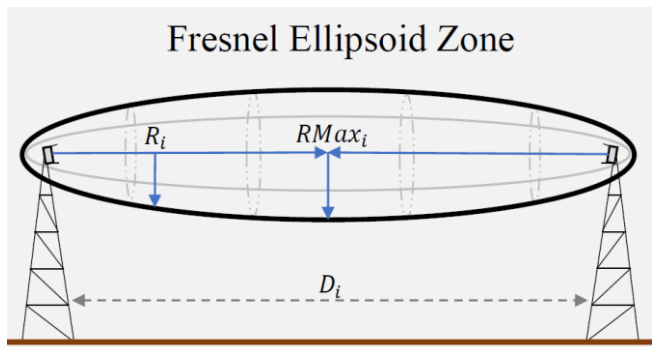

**Figure 4** Fresnel ellipsoid zone

This is because to ensure signal can be transmitted between two wireless backhaul antennas, the Fresnel Zone needs to be kept free of interference from buildings and other structures. In rural and remote areas, the link path needs to be kept free of the tree canopy. Thus, to find the *maximum* radius of the Fresnel Ellipsoid Zone ($RMax_i$), between the two antennas in the $i$th wireless backhaul, we use (4) based on both the measured distance ($D_i$) and desired carrier frequency ($F_i$):

$$RMax_i = 8.66 \cdot \sqrt{\frac{D_i}{F_i}} \qquad (4)$$

Subsequently, for different backhaul frequency ranges, the Fresnel clearance can be plotted against the potential distance between the antennas, as reported in Fig. 3 (B). The results indicate that both lower frequencies and longer link distances require higher Fresnel clearances. For example, the median Fresnel clearance values below 10 km are approximately 5-6 meters depending on the frequency. But these values increase to 9-13 meters between 10-25 km, and then to 12-18 meters for 25-45 km. Generally, the distance has a larger impact on the required Fresnel clearance, compared to the frequencies used here. The estimated values prove to be highly useful for approximating tower heights and therefore necessary civil engineering costs. Next, the strategies to be tested will be discussed.

### B. STRATEGIES

As the approach is based on applying backhaul technologies within a techno-economic framework, the strategies tested are guided by the scientific detail reported both in this manuscript and in previous field studies [24], [25]. The aim is to carry out system-wide national assessment of CLOS and hybrid CLOS-NLOS backhaul strategies, utilizing both population density and terrain irregularity for evaluation in different locations.

Firstly, Strategy 1 aims to deploy only CLOS links, using knowledge of all settlements which require connecting back into a major settlement using wireless means. Secondly, Strategy 2 takes a similar approach but instead utilizes a hybrid deployment of both CLOS and diffractive NLOS technologies. This strategy will always preference CLOS links but attempt to use a diffractive NLOS approach as a last resort, enabling a single 'hop' over a 'knife-edge' diffraction point.

The strategies are implemented in combination with a set of least-cost infrastructure design algorithms, which are originally developed for this assessment and applied within a geospatially-explicitly simulation framework. Least-cost infrastructure algorithms are a state-of-the-art way to design and test telecom and energy network strategies [90], [91].

### B. SPATIAL PROCESSING

Prior to testing the strategies identified, a set of preprocessing steps are first required to manipulate various data layers into a common format. This begins with identifying settlements, which in this analysis are defined as any area having more than 50 inhabitants per km², with an overall total of 100 inhabitants. This is a common methodological step in telecom and energy strategy assessment [92]. Then it is necessary to find the nearest major settlement which has over 20,000 inhabitants [93], and therefore strong economics for existing digital connectivity, including a fiber PoP which can help route traffic to the Internet.

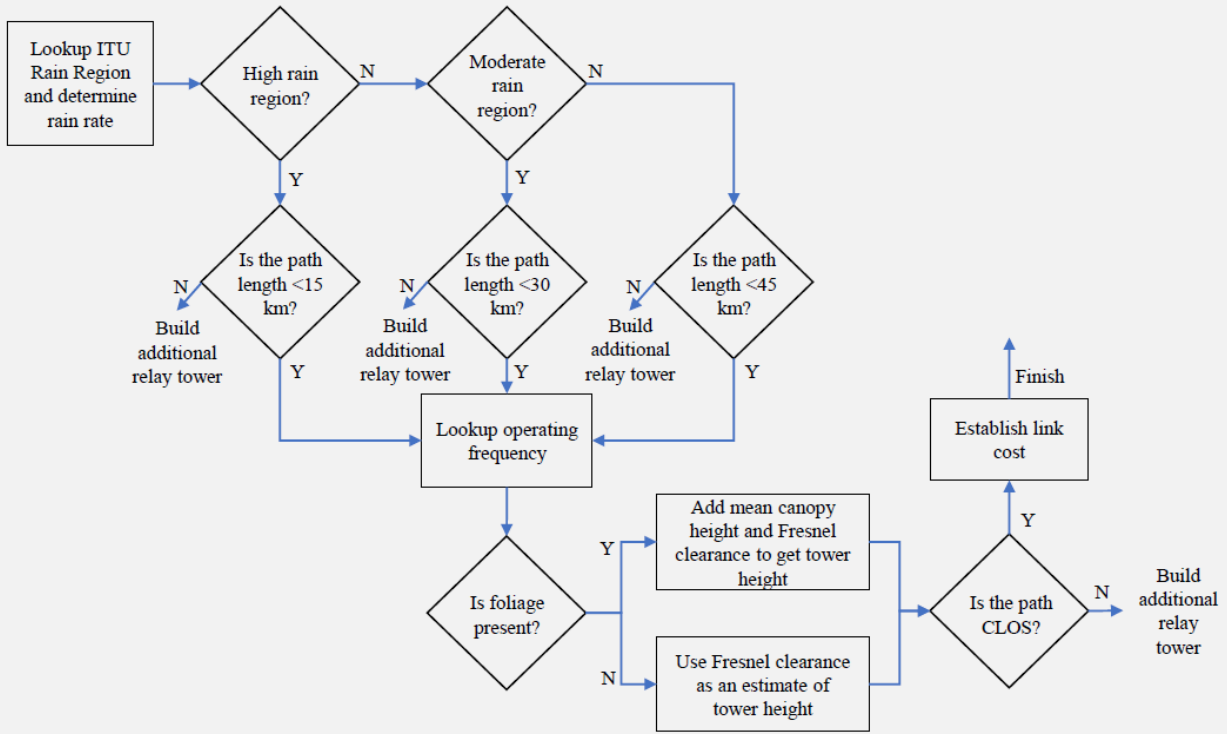

**Figure 5** Flow diagram for CLOS

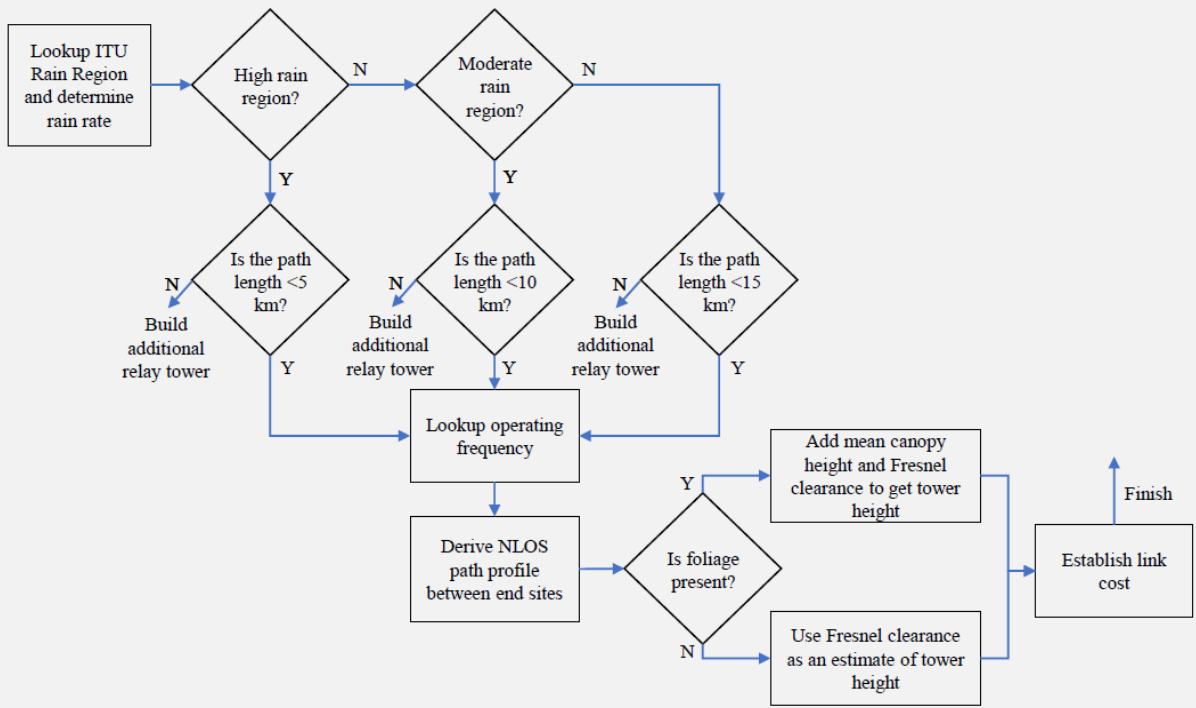

**Figure 6** Flow diagram for NLOS

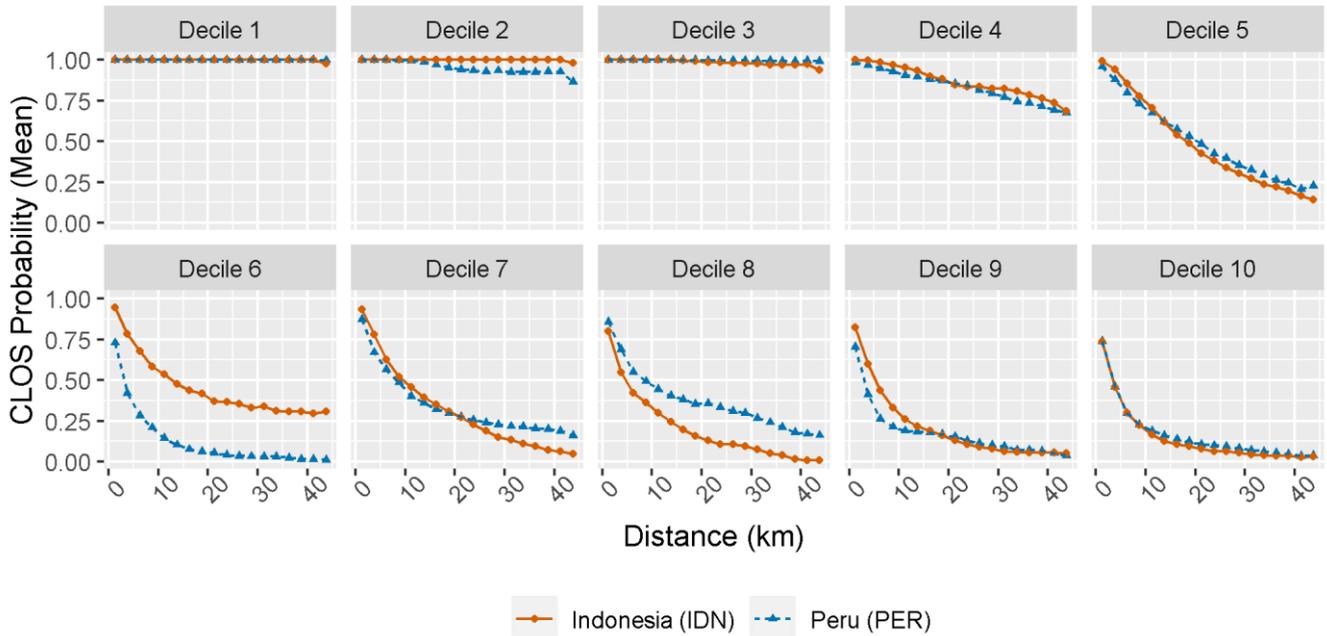

**Figure 7** Visualization of LOS sampling results

These two steps provide the nodes to connect (small settlements) and the nodes to route traffic to (major settlements). Spatial autocorrelation issues can arise due to artificial statistical boundaries [94], [95], therefore in such spatial statistical problems it is wise to generate bespoke area boundaries which best represent the problem being modeled. This is a common issue for strategic assessment of telecom and energy network options [96], [97]. Consequently, a set of 'modeling regions' are defined which consist of merging local areas to ensure each has a least one major settlement to route data traffic to. This is achieved by taking the straight-line route from each local settlement to a major settlement, creating a union of any statistical areas which these lines intersect with.

Once defined, it is possible to connect all settlements within each modeling region by fitting a minimum spanning tree based on Dijkstra's algorithm, which is a common approach for least-cost telecom network design and routing [98]–[100]. Thus, the shortest potential path is estimated to create the least-cost network structure that connects all nodes in each modeling region.

A final preprocessing step estimates CLOS probability for various terrain types over different distances. Firstly, as a preparatory step, any target country is divided into 50x50 km$^2$ grid tiles, with each tile therefore covering a 2,500 km$^2$ area. Based on the Longley-Rice Irregular Terrain Model, a terrain irregularity parameter is extracted for each tile [101].

To obtain this metric, elevation data are extracted from a Digital Elevation Model enabling the inter-decile range of terrain elevation values to be obtained. The Longley-Rice

Irregular Terrain Model is a classic radio propagation model and capable of estimating signal propagation effects resulting from irregular terrain [102], [103]. Once this parameter has been estimated for the whole country grid, the dataset is split into deciles and a sampling area from each decile is randomly selected. This area is then converted to a 2.5x2.5 km$^2$ grid to enable a random point to be selected every 6.25 km$^2$.

A viewshed is then computed using Python via the Geographic Resources Analysis Support System (GRASS) package to estimate the LOS for every sample point, to every other point within a maximum of 45 km$^2$, based on mean tower heights of 30 meters (a standard height for cellular network dimensioning [104]). This exercise produces a lookup table of the LOS probability given a particular terrain decile over a specific distance, as illustrated for the countries assessed in Fig. 7.

### C. MODELING PROCESS
In this section the modeling process is outlined, including the originally developed flow diagrams each strategy follows, which are also shown visually for CLOS in Fig. 5 and NLOS in Fig. 6. As CLOS is the current business-as-usual case for wireless backhaul deployment, the approach will be familiar to wireless communications researchers. In contrast, the NLOS flow diagram process illustrated in Fig. 6 is newly specified from field empirical results for Jaén, Peru [25]. In this section, the model process is discussed in general terms. Later, the specific required datasets for the model are reported in Section V, which focuses on the application of the model.

Firstly, for each region of a country under evaluation an ITU determined rain region is allocated to account for air moisture effects on propagation [105]. For a high, moderate, or low rain region the maximum CLOS paths are conservatively defined as 15, 30 and 45 kilometers, respectively. Although these significantly downgrade the distances reported in Fig. 3, the approach adopted is a cautious one, based on the premise that it is worse to underestimate cost at the initial planning stage, than overestimate. For example, if costs savings are made when more detailed RF planning is carried out at a later stage, this is more acceptable than finding out the initial project thought to be viable, is no longer possible, after investing further resources to no avail.

The three main backhaul operating frequencies already used in the method are considered and allocated based on the link distance. We know from Fig. 3 (A) that lower frequencies will be preferable for longer distances, and vice versa. Thus, CLOS links under 10 kilometers use the highest frequency of 18 GHz, intermediate link distances (10-25 km) use 15 GHz, and long-range link distances (25-45 km) use 8 GHz. Whereas NLOS links under 5 kilometers use the highest frequency of 18 GHz, intermediate link distances (10 km) use 15 GHz, and long-range link distances (15 km) use 8 GHz.

Using the preprocessed spatial routing geometries, it is then possible to iterate through each routing path between settlements using a two-step viewshed computation approach. For each preprocessed routing path, a viewshed is explicitly carried out, and if a new relay tower needs to be built (as per Fig. 5 and Fig. 6), the LOS of this new asset is based on a probability derived from the lookup table for Fig. 7. This approach balances the first-order precision of computing a specific viewshed, with the computational speed of using a probability for any second-order links that need building. This part of the method is similar to other studies in the literature which utilize interactive testing of line-of-sight [28].

Based on the literature, a mean tower height of 30 meters is used to run the viewshed processing [104], to establish line-of-sight. If the result of the viewshed processing for a particular route is CLOS, the flowchart in Fig. 5 is used for the network design.

To determine the required tower height, the presence of foliage is then assessed via remote sensing techniques using globally available satellite data layers. From existing studies in the literature, we know that areas with foliage cover have a higher probability of requiring additional Fresnel clearance [106], [107] (for the remote rural area focus of this assessment).

If foliage is present (>20%), then the mean height of the tree canopy is added to the Fresnel clearance values obtained from Fig. 3 (B) and reported for different confidence levels in Table 2. If no foliage is present, the Fresnel clearance is used for the tower height.

TABLE 1
FRESNEL CLEARANCE HEIGHTS FROM GROUND

| Link Distance (Km) | Link Frequency (GHz) | Fresnel Clearance (M) by Confidence Level | | |
|---|---|---|---|---|
| | | 50% | 90% | 99% |
| <10 | 6 to 8 | 6.7 | 9.4 | 10.4 |
| <10 | 11 to 15 | 6.1 | 7.5 | 8.1 |
| <10 | 15 to 18 | 4.7 | 6.6 | 6.9 |
| 10-25 | 6 to 8 | 13.2 | 15.7 | 17.0 |
| 10-25 | 11 to 15 | 10.1 | 11.9 | 12.6 |
| 10-25 | 15 to 18 | 9.0 | 10.2 | 11.0 |
| 25-45 | 6 to 8 | 18.7 | 21.0 | 22.4 |
| 25-45 | 11 to 15 | 14.2 | 16.1 | 17.3 |
| 25-45 | 15 to 18 | 12.1 | 13.3 | 14.3 |

Finally, antennas cannot be mounted directly on top of erected towers due to high wind exposure. Indeed, antennas need to be secured directly to the side of the metal tower structure in multiple places to provide strong wind protection. Therefore, a mean additional height of one meter is added to the total estimated tower height to account for this antenna mounting, based on the antenna sizes reported in Table 3.

Although two separate strategies are to be tested, the key differentiator is that much smaller maximum path distances are used for NLOS to reflect the more challenging terrain conditions. Link distances are as short as 5 kilometers in high rain regions, as a higher fade margin is used to represent greater uncertainty and therefore more challenging QoS conditions when using an NLOS link [12]. In both cases, if the link is over the maximum distance, a relay tower is required, which is the key cost driver and the reason why CLOS approaches have poor cost efficiency in hard-to-reach areas.

As a high-level planning method is adopted here, we assume any diffractive NLOS link crosses a single diffracting obstacle and that the diffracting angle is shallow (between 177-180 degrees). Indeed, existing analysis has demonstrated that if the diffraction angle is kept to 3 degrees, and the diffraction loss to 25 dB, then standard link planning tools can make fairly accurate predictions of signal level (with a 10 dB additional margin to correct for prediction error, irregularity in the shape of the diffracting object, foliage changes through the seasons, and terrain data errors) [25]. Indeed, the live network tests for link and cluster designs in Jaén, Peru, show that even with such restrictions, diffractive NLOS still delivers a significant new capability. Even when much shorter maximum path lengths are applied here for diffractive NLOS assessment, this remains a feasible expectation, providing sufficient detail for initial planning for prioritizing resource allocation.

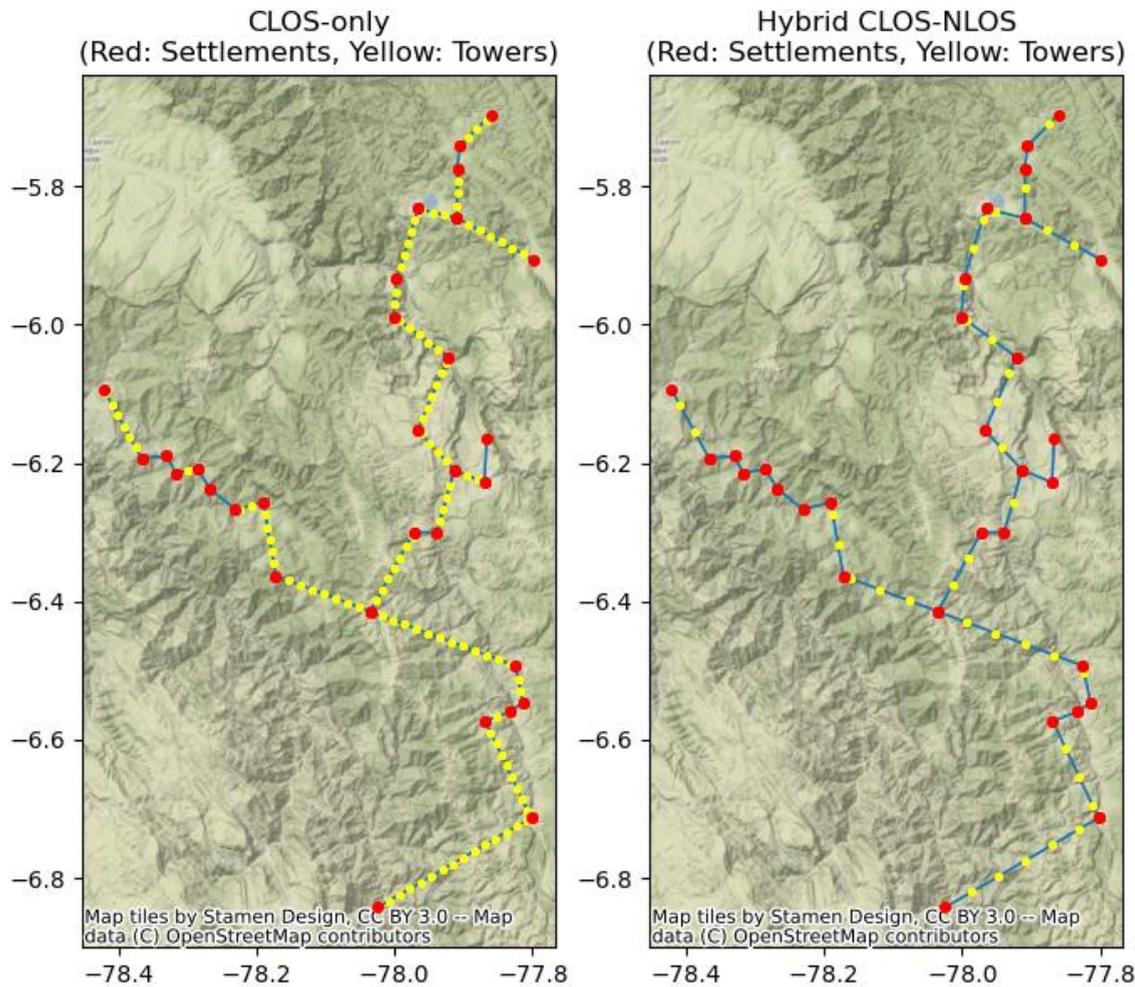

**Figure 8** Example visualization of network designs

TABLE 3
COST ITEMS

| Equipment Item | Caveats | Cost (USD) | Source |
|---|---|---|---|
| Two PtP radios (all-ODU, high power, licensed bands) | - | 6,000 | [108] |
| Two 0.6 m PtP parabolic antennas | Link distances: <10 km | 1,200 | [109] |
| Two 0.9 m PtP parabolic antennas | Link distances: 10 − 20 km | 2,200 | [110] |
| Two 1.2 m PtP parabolic antennas | Link distances: 20 − 30 km | 3,600 | [111] |
| Two 1.8 m PtP parabolic antennas | Link distances: 30 - 45 km | 4,460 | [112] |
| Tower cost per 10 m in height (Materials, construction, transportation etc.) | - | 10,000 | [113] |
| Network planning, site acquisition and installation | - | 8,700 | [114] |
| Single site PV + battery power system | - | 12,000 | [115] |

## C. EQUIPMENT COSTS

Generally, network planners attempt to optimize tower construction costs, preferencing where possible for the smallest tower height which still achieves certain quality of service objectives. Generally, a tower can be erected relatively cheaply (e.g., 30 meter) with a freestanding structure using no guide wires to ensure full clearance of any present tree canopy. In some cases, larger assets may be required.

The investment cost ($Cost_r$) for the modeling region ($r$) is the summation of all cost inputs for the $i$th wireless link, as defined in (5):

$$Cost_r = \sum_{i=0}^{n} r_i + a_i + t_i + p_i + e_i \qquad (5)$$

To connect each settlement into the wireless backhaul network, a pair of Point-to-Point (PtP) radios ($r_i$) and parabolic backhaul antennas ($a_i$) are required, along with the civil engineering costs of building a tower ($t_i$), site planning ($p_i$), and provision of a suitable power system ($e_i$). Full item costs are reported in Table 3 based on indicative equipment prices for 2021, with these values comparing well to other studies in the IEEE literature [116], published by GSMA [117], or other industry consultants [114].

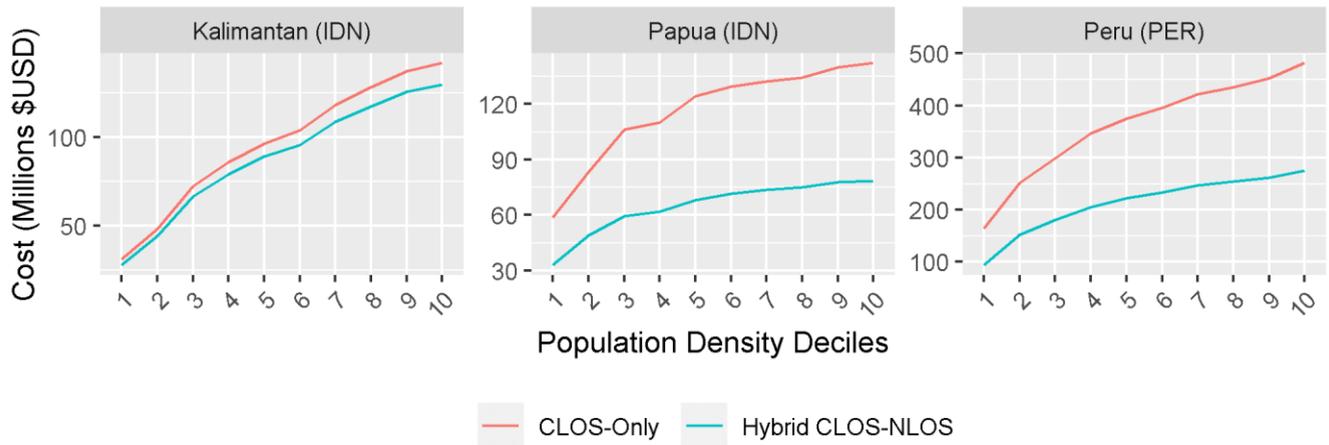

### (A) Cumulative Investment Cost by Population Density Deciles
Deciles labelled from the highest density to the lowest (1-10)

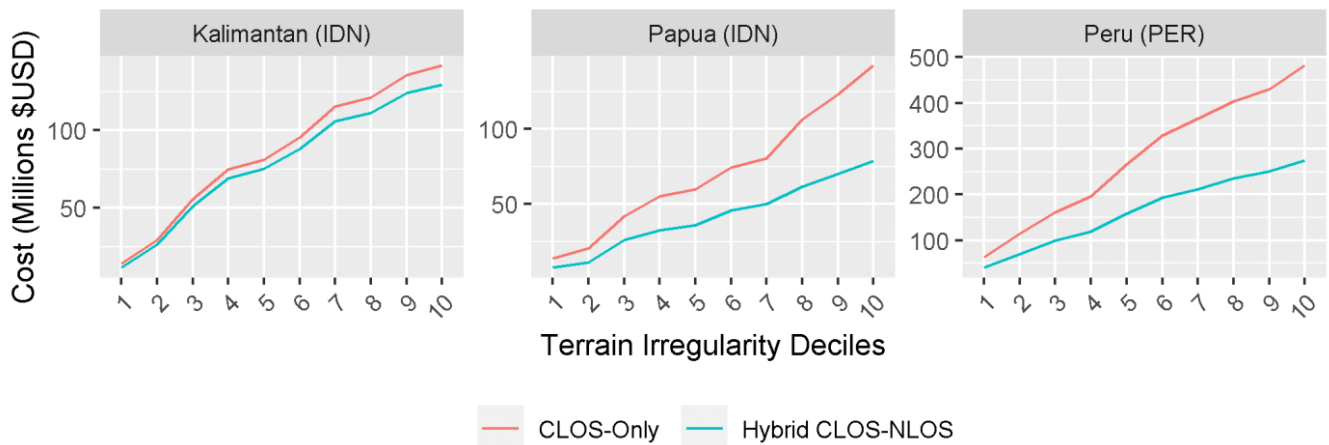

### (B) Cumulative Investment Cost by Terrain Irregularity Deciles
Deciles labelled from the lowest terrain irregularity to the highest (1-10)

**Figure 9** Cumulative cost results reported by population density and terrain irregularity

As well as scaling the cost based on the required tower height needed to clear the tree canopy (with this scaling being driven by the material transportation cost), the link distance also affects the size of the required parabolic antennas. These range from 0.6 meters for under 10 kilometers, up to 1.8 meters for links at the maximum distance of 45 kilometers.

## V. APPLICATION

Two countries are used to apply the method develop here, including Indonesia and Peru. In Indonesia, we focus on two of the largest islands (Kalimantan and Papua) and in Peru we assess the whole country. Kalimantan has a population of approximately 17 million inhabitants and is growing in importance due to the capital of Indonesia being relocated there from Jakarta. Internet access already exists in the major urban settlements but is scarcely available in rural areas. Papua has approximately 3.4 million inhabitants with much of the island unconnected by broadband connectivity at all. In contrast, Peru has an approximate population of 33 million inhabitants and while there is broadband access in all major urban areas, connectivity is progressively poorer as the country stretches eastwards from the Pacific coastline. This is driven by the challenging terrain in the Andes mountain range, as well as the thick rainforest tree canopy in the Peruvian Amazonia region.

Regional boundaries are obtained from the Database of Global Administrative Areas (GADM) [118]. Level 2 regions produce 90 statistical areas for Indonesia (Kalimantan and Papua) and 195 Peru. After preprocessing, this results in 58 modeling regions for the two Indonesian islands and 95 for Peru. Next, a global 1 km$^2$ population layer is obtained from WorldPop [119], [120] and used to derive the settlement point layer for network routing to major settlements [25].

Various remote sensing layers are integrated because they provide globally available data inputs for establishing terrain and vegetation cover. This is important for model scalability, as future users of the open-source e3nb codebase [70] may wish to apply the capability to new countries (thus, minimal code changes would be required).

The Global Multi-resolution Terrain Elevation Data (GMTED) (2010) is used as the Digital Elevation Model, available from the US Geological Survey as a raster data layer (.tif) [121]. To identify vegetation presence, the NASA Moderate Resolution Imaging Spectroradiometer (MODIS) data layer is obtained which estimates the proportion of vegetation cover from high-resolution imagery (.tif) [122]. To estimate the mean canopy height of tree foliage in a modeling region, results are utilized from the LiDAR-derived Geoscience Laser Altimeter System (GLAS) Global Estimates of Forest Canopy Height [123].

After running the preprocessing steps for each country to obtain the LOS lookup, a set of probabilities can be used for CLOS availability over different distances in each country, given ten different terrain irregularity types. Decile 1 has the least terrain irregularity, and Decile 10 the most. The results are reported in Fig. 7.

The complement of the CLOS probability is the NLOS probability. Therefore, in Decile 1 the simulation results indicate almost a 100% probability of CLOS (thus, a 0% probability of NLOS), whereas in Decile 10 there is a much smaller chance of a CLOS link being feasible, depending on the distance. Generally, the LOS is quite similar in both countries for the lower and upper deciles, although in the middle deciles Peru has more irregular terrain. The results of the applied method can now be reported.

## VII. RESULTS
A key contribution of this approach is in helping to prioritize areas of future investment, thanks to an explicitly spatial methodology. As an example of the different automated network designs produced for the techno-economic assessment, a regional illustration for a mountainous rural area in Peru is provided in Fig. 8. These automated designed build on the knowledge gained from the manual network designs produced in [25]. Identified settlements are plotted in red, with required towers plotted in yellow, and the least-cost network design connecting all settlements in blue.

The modeling region contains a city of approximately 30,000 inhabitants called Chachapoyas, surrounded by many smaller settlements ranging from 5,000 down to only 250 inhabitants. While the main city has cellular connectivity, surrounding settlements have very little access. Challenging deployment conditions result from low average revenue per user and modest adoption on the demand-side, combined with highly irregular terrain affecting the supply-side, as traditional CLOS wireless connectivity becomes very expensive as many more towers may need to be built.

In the strategy which utilizes only CLOS links, an estimated 146 towers are required to build a regional wireless backhaul network which can connect all settlements in Fig. 8, based on the conservative planning criteria which the high-level evaluation method defined. In contrast, for a hybrid strategy which can deploy a mixture of CLOS and diffractive NLOS links, an estimated 69 towers are required. This leads to significant investment ramifications for unconnected areas.

The cumulative cost of each strategy is reported for both strategies for the full assessments of Kalimantan and Papua in Indonesia, and Peru. Fig. 9 illustrates the cumulative cost results by country reported using the population density (A) and terrain irregularity (B) deciles.

The sets of plots in Fig. 9 allow the y-axis to vary, enabling the main differences between the two different strategies to be visualized more effectively. In Kalimantan, Indonesia, the cost of serving the settlements identified reaches approximately $142 million using CLOS, compared to $129 million with a hybrid approach, equating to a 9 percent saving. In contrast, Papua in Indonesia, produced a CLOS estimate for serving the identified settlements of $142 million, relative to $79 million with a hybrid approach, leading to a 45 percent cost saving. Finally, in Peru, the national cost of using CLOS to connect all identified settlements is approximately $482 million, considerably more than the $276 million estimated using a hybrid strategy, delivering a 43 percent saving.

These results demonstrate that cost savings produced by deploying diffractive NLOS wireless backhaul links are generally correlated with both population density and terrain irregularity. For example, in Fig. 9 Kalimantan in Indonesia produces a similar cumulative cost curve when ranked by each factor. Moreover, when comparing the two deployment strategies, in Papua , Indonesia, CLOS-only approaches rise rapidly in cost in the final three deciles. This contrasts with a much more modest increase when using a hybrid approach with diffractive NLOS links, as fewer towers are required to cover the same distance.

Following the same logic, Fig. 10 illustrates the aggregate cost results per decile when ranked by either population density or terrain irregularity. The cost saving for each decile between the strategies (CLOS versus hybrid CLOS-NLOS) can be compared, with the hybrid deployment approach being cheaper in most deciles. These savings range from 4-13% and 27-57% for Kalimantan and Papua in Indonesia, respectively, and 33-53% for Peru.

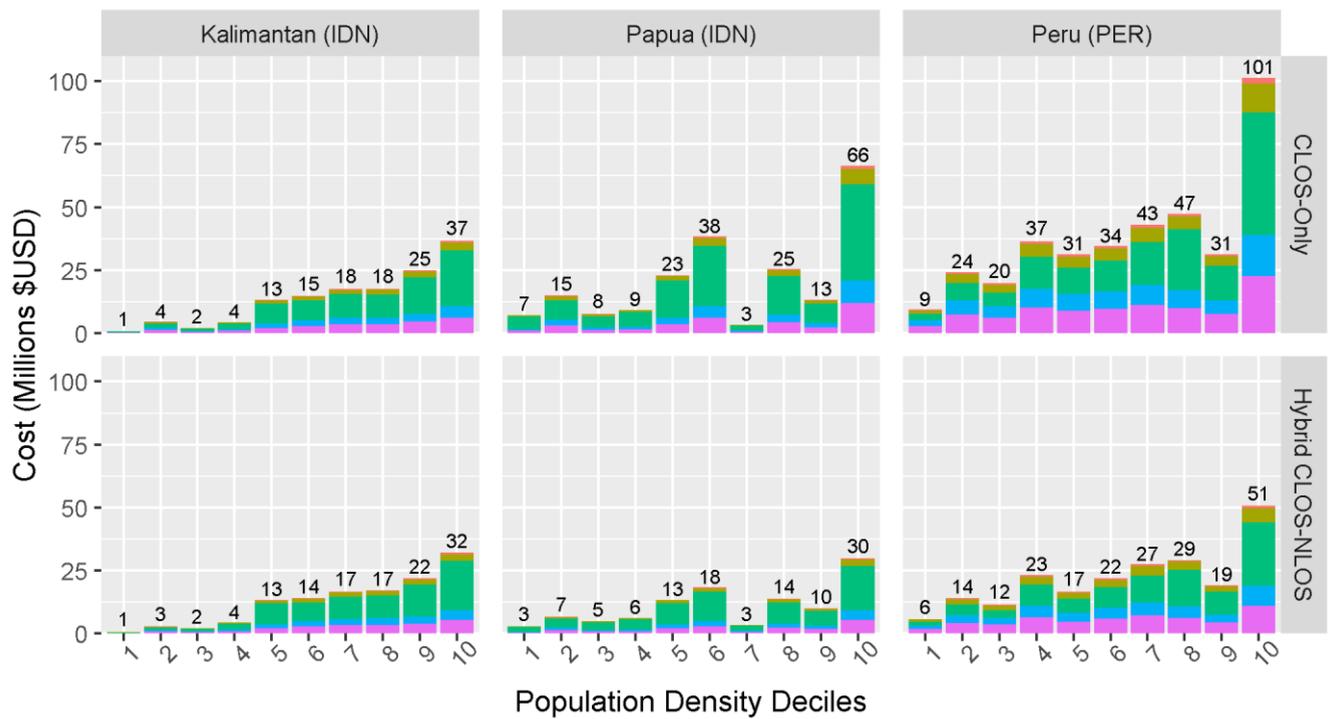

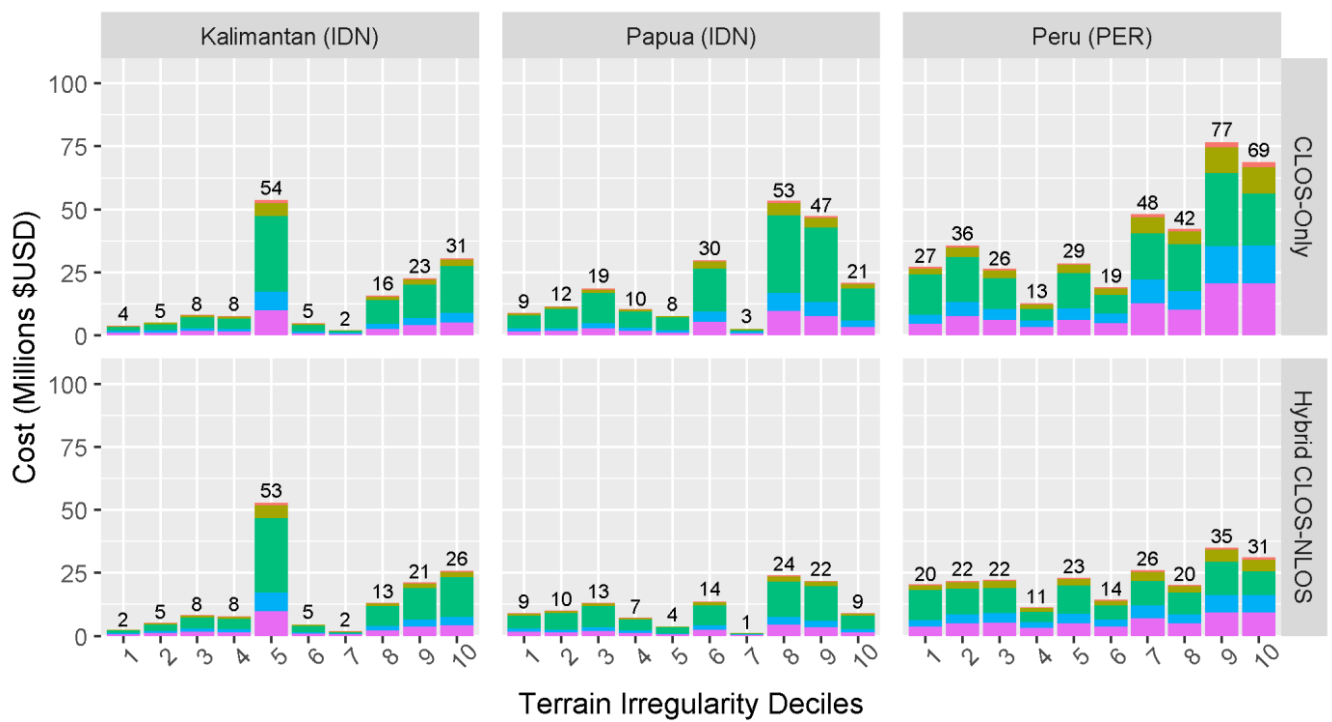

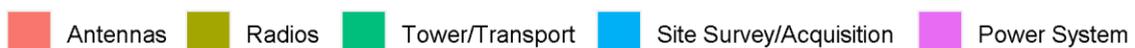

**Figure 10 Aggregate cost results reported by population density and terrain irregularity**

Fig. 10 also presents the composition of this cost based on the required investment into antennas, radios, tower construction and transportation, site survey/acquisition, and the site power system. Broadly speaking, the tower construction and transportation of materials was the largest overall capital expenditure cost, with the contribution of this factor elevated in those areas with substantial tree canopy heights, such as in Papua, Indonesia. This results from the need for a larger Fresnel clearance zone. The only caveat to the results in Fig. 10 is that the modeling regions used vary between 90-160 statistical units per country. Future research may want to use more granular spatial boundaries which would provide a smoother decile cost curve.

The results highlight the challenge of using CLOS approaches to serve settlements in mountainous areas. Often these settlements are in the bottom of valleys, making them particularly hard to connect. However, in certain circumstances, diffractive NLOS can make this possible without building additional towers, leading to substantial cost savings, even if the wireless links must be under distances as short as 5 kilometers. The contributions of this paper can now be revisited in the following conclusionary section.

## VIII. Discussion of limitations
All research has limitations. Therefore, it is important to identify shortcomings as they serve as important areas of future research. To begin, the least-cost routing paths were exogenously fixed within this assessment to improve the efficiency of the codebase. This means that the network designs do not provide an optimal solution due to the complexity in undertaking such a high-dimensional task. Future research needs to take advantage of greater computing resources to explore the ramifications of these simplifications, especially as we know spatio-temporal factors are very important when modeling wireless networks [124], for example, for antenna locations [125], spectrum sensing [126], channel simulation [127] and NLOS [128]. The future use of stochastic spatial simulation methods could help actively explore potential tower placement along a routing path corridor. These advancements may help to refine the cost estimates used here.

Moreover, this assessment work focused only on including terrain irregularity effects via a Digital Elevation Model but did not include surface blockage effects (e.g., via a Digital Surface Model). Future research could better utilize 3D building height information, to address this area of uncertainty in the network planning assessment process.

Given the shifting dynamics of the telecoms industry, it is also worth considering how the method framework could be adapted for a post-5G/6G world. This is especially given the increasing role machine learning is likely to play in network optimization. Firstly, global connectivity has already started to become a use case for 6G [85], [129], based on the idea that every location should be connected wirelessly, with a much larger role for satellite constellations and Unmanned Aerial Vehicles (UAVs). Therefore, this method could be adopted to account for the increased need for 3D network assessment tools for these types of network architectures. Secondly, one of the main computer languages used for many machine learning packages is Python, therefore the open-source e3nb [70] repository could be adapted to explore new techno-economic optimization methods using machine learning techniques.

The assessment framework presented here provides a starting point for detailed 3D wireless techno-economic backhaul planning, but such endeavors will require close collaboration with network operators. Traditionally, techno-economic assessment does not consider 3D effects, focusing only on two-dimensional distance impacts. Research to further develop the e3nb codebase should focus on working in close partnership with operators to gain input on the classification and prioritization of settlements, and 3D network design process for different wireless backhaul technologies.

## IX. CONCLUSION
In this paper several key scientific contributions have been made to the literature.

Firstly, an engineering-economic evaluation of different wireless backhaul strategies has been undertaken. This assessment is based on (i) a traditional CLOS approach versus (ii) a hybrid approach using CLOS where possible, and diffractive NLOS as a backup in challenging terrain situations. Given the more recent emergence of diffractive NLOS wireless backhaul research, this is (to our knowledge) one of the first techno-economic assessment of this technology for backhaul links. Therefore, the method and results are of substantial relevance to operators, governments and other agencies interested in providing equitable universal broadband connectivity for all of society. The fact that the results estimate a cost saving of approximately 9-45 percent is certainly encouraging. Indeed, spare capital from this efficiency gain can be reallocated into connecting more unconnected users, helping to reduce the billions of people globally yet to access a decent broadband service.

Secondly, from a more theoretical modeling framework perspective, the method developed here focused on advancing initial high-level assessment techniques in a spatially explicit way. This complements the existing wireless backhaul literature, where most engineering papers have understandably focused on assessing QoS at the cluster or link level. They have also not included techno-economic considerations. Additionally, most techno-economic models published in the engineering literature usually provide a relatively simplistic, sometimes spreadsheet-based, approach to cost assessment. In contrast, this paper utilized a combination of remote sensing and viewshed geospatial techniques. Therefore, balancing detailed estimations using viewsheds, with the computational efficiency of using probability-based line-of-sight for different terrain types. The techno-economic assessment framework presented can be used to help prioritize regions for future investment for broadband connectivity.

ACKNOWLEDGMENT
The authors would like to thank industry partners from the mobile sector who have supported the research. This includes colleagues from Internet para Todos such as Renan Ruiz Moreno, Juan Manuel Perez, and Manuel Garcia Lopez. Colleagues from Mayutel including Omar Tupayachi Calderon, Carlos Araujo Castro and Diego Ramos Chang. As well as colleagues from TeleworX including José Huarcaya, Diego Mendoza, Henry Claver and Miranda Flores. The authors would like to thank Facebook Connectivity for enabling the research to take place via an open-science grant.

REFERENCES

[1] United Nations, 'Infrastructure and Industrialization', *United Nations Sustainable Development*, 2021. https://www.un.org/sustainabledevelopment/infrastructure-industrialization/ (accessed Nov. 08, 2021).

[2] UNCTAD, 'UNCTAD DGFF2016 SDG Goal 9.c - Access to ICT', *UNCTAD DGFF 2016*, 2021. https://stats.unctad.org/Dgff2016/prosperity/goal9/target_9_c.html (accessed Nov. 08, 2021).

[3] A. López-Vargas, A. Ledezma, J. Bott, and A. Sanchis, 'IoT for Global Development to Achieve the United Nations Sustainable Development Goals: The New Scenario After the COVID-19 Pandemic', *IEEE Access*, vol. 9, pp. 124711–124726, 2021, doi: 10.1109/ACCESS.2021.3109338.

[4] L. Chiaraviglio *et al.*, 'Bringing 5G into Rural and Low-Income Areas: Is It Feasible?', *IEEE Communications Standards Magazine*, vol. 1, no. 3, pp. 50–57, Sep. 2017, doi: 10.1109/MCOMSTD.2017.1700023.

[5] R. Borralho, A. Mohamed, A. Quddus, P. Vieira, and R. Tafazolli, 'A Survey on Coverage Enhancement in Cellular Networks: Challenges and Solutions for Future Deployments', *IEEE Communications Surveys Tutorials*, pp. 1–1, 2021, doi: 10.1109/COMST.2021.3053464.

[6] Y. Li, A. Cai, G. Qiao, L. Shi, S. K. Bose, and G. Shen, 'Multi-Objective Topology Planning for Microwave-Based Wireless Backhaul Networks', *IEEE Access*, vol. 4, pp. 5742–5754, 2016, doi: 10.1109/ACCESS.2016.2581187.

[7] C. Madapatha *et al.*, 'On Integrated Access and Backhaul Networks: Current Status and Potentials', *IEEE Open Journal of the Communications Society*, vol. 1, pp. 1374–1389, 2020, doi: 10.1109/OJCOMS.2020.3022529.

[8] C. Saha and H. S. Dhillon, 'Millimeter Wave Integrated Access and Backhaul in 5G: Performance Analysis and Design Insights', *IEEE Journal on Selected Areas in Communications*, vol. 37, no. 12,

pp. 2669–2684, Dec. 2019, doi: 10.1109/JSAC.2019.2947997.

[9] E. Zeydan, O. Dedeoglu, and Y. Turk, 'Performance monitoring and evaluation of FTTx networks for 5G backhauling', *Telecommun Syst*, vol. 77, no. 2, pp. 399–412, Jun. 2021, doi: 10.1007/s11235-021-00767-0.

[10] T. Sharma, A. Chehri, and P. Fortier, 'Review of optical and wireless backhaul networks and emerging trends of next generation 5G and 6G technologies', *Transactions on Emerging Telecommunications Technologies*, vol. 32, no. 3, p. e4155, 2021, doi: 10.1002/ett.4155.

[11] I. del Portillo, B. G. Cameron, and E. F. Crawley, 'A technical comparison of three low earth orbit satellite constellation systems to provide global broadband', *Acta Astronautica*, vol. 159, pp. 123–135, Jun. 2019, doi: 10.1016/j.actaastro.2019.03.040.

[12] N. J. Frigo, P. P. Iannone, and K. C. Reichmann, 'A view of fiber to the home economics', *IEEE communications Magazine*, vol. 42, no. 8, pp. S16–S23, 2004.

[13] O. B. Osoro and E. J. Oughton, 'A Techno-Economic Framework for Satellite Networks Applied to Low Earth Orbit Constellations: Assessing Starlink, OneWeb and Kuiper', *IEEE Access*, vol. 9, pp. 141611–141625, 2021, doi: 10.1109/ACCESS.2021.3119634.

[14] I. del Portillo, S. Eiskowitz, E. F. Crawley, and B. G. Cameron, 'Connecting the other half: Exploring options for the 50% of the population unconnected to the internet', *Telecommunications Policy*, vol. 45, no. 3, p. 102092, Apr. 2021, doi: 10.1016/j.telpol.2020.102092.

[15] M. M. Ahamed and S. Faruque, '5G Backhaul: Requirements, Challenges, and Emerging Technologies', *IntechOpen*, 2018, doi: 10.5772/intechopen.78615.

[16] W. Guo and T. O'Farrell, 'Relay Deployment in Cellular Networks: Planning and Optimization', *IEEE J. Select. Areas Commun.*, vol. 31, no. 8, pp. 1597–1606, Aug. 2013, doi: 10.1109/JSAC.2013.130821.

[17] M. K. Shukla, S. Yadav, and N. Purohit, 'Secure Transmission in Cellular Multiuser Two-Way Amplify-and-Forward Relay Networks', *IEEE Transactions on Vehicular Technology*, vol. 67, no. 12, pp. 11886–11899, Dec. 2018, doi: 10.1109/TVT.2018.2877133.

[18] Z. Fang, Y. Wu, Y. Lu, J. Hu, T. Peng, and J. Ye, 'Simultaneous Wireless Information and Power Transfer in Cellular Two-Way Relay Networks With Massive MIMO', *IEEE Access*, vol. 6, pp. 29262–29270, 2018, doi: 10.1109/ACCESS.2018.2834534.

[19] Z. Liao, J. Liang, and C. Feng, 'Mobile relay deployment in multihop relay networks', *Computer*

*Communications*, vol. 112, pp. 14–21, Nov. 2017, doi: 10.1016/j.comcom.2017.07.008.

[20] B. Kim and T. Kim, 'Relay Positioning for Load-Balancing and Throughput Enhancement in Dual-Hop Relay Networks', *Sensors*, vol. 21, no. 5, Art. no. 5, Jan. 2021, doi: 10.3390/s21051914.

[21] Z. Fang, W. Ni, F. Liang, P. Shao, and Y. Wu, 'Massive MIMO for Full-Duplex Cellular Two-Way Relay Network: A Spectral Efficiency Study', *IEEE Access*, vol. 5, pp. 23288–23298, 2017, doi: 10.1109/ACCESS.2017.2766079.

[22] J. Xu, Y. Zou, S. Gong, L. Gao, D. Niyato, and W. Cheng, 'Robust Transmissions in Wireless-Powered Multi-Relay Networks With Chance Interference Constraints', *IEEE Transactions on Communications*, vol. 67, no. 2, pp. 973–987, Feb. 2019, doi: 10.1109/TCOMM.2018.2877466.

[23] S. Gong, S. X. Wu, A. M. So, and X. Huang, 'Distributionally Robust Collaborative Beamforming in D2D Relay Networks With Interference Constraints', *IEEE Transactions on Wireless Communications*, vol. 16, no. 8, pp. 5048–5060, Aug. 2017, doi: 10.1109/TWC.2017.2705062.

[24] J. Kusuma, E. Boch, and P. Liddell, 'Diffractive NLOS microwave backhaul for rural connectivity', Telecom Infra Project, Massachusetts, 2021. [Online]. Available: https://cdn.brandfolder.io/D8DI15S7/as/4pbj354s4zff94vs9xgmx3s/TIP-NaaS_White_Paper_Diffractive_NLOS_Microwave_Backhaul_for_Rural_Connectivity_January_2021.pdf

[25] J. Kusuma and E. Boch, 'Improving Rural Connectivity Coverage using Diffractive Non-Line of Sight (NLOS) Wireless Backhaul', Malaysia, 2021. [Online]. Available: https://research.fb.com/wp-content/uploads/2021/01/Improving-Rural-Connectivity-Coverage-using-Diffractive-Non-Line-of-Sight-NLOS-Wireless-Backhaul.pdf

[26] Internet Para Todos, 'Internet Para Todos', *Internet Para Todos*, 2021. https://www.iparatodos.com.ar/ (accessed Nov. 08, 2021).

[27] Mayu telecomunicaciones, 'Mayu telecomunicaciones', *Mayu telecomunicaciones*, 2021. http://mayutel.com/inicio.html (accessed Nov. 08, 2021).

[28] P. E. Brown, K. Czapiga, A. Jotshi, Y. Kanza, and V. Kounev, 'Interactive Testing of Line-of-Sight and Fresnel Zone Clearance for Planning Microwave Backhaul Links and 5G Networks', in *Proceedings of the 28th International Conference on Advances in Geographic Information Systems*, New York, NY, USA, Nov. 2020, pp. 143–146. doi: 10.1145/3397536.3422332.

[29] P. E. Brown, K. Czapiga, A. Jotshi, Y. Kanza, V. Kounev, and P. Suresh, 'Large-Scale Geospatial Planning of Wireless Backhaul Links', p. 4, 2020.

[30] J. Baek and Y. Choi, 'Comparison of Communication Viewsheds Derived from High-Resolution Digital Surface Models Using Line-of-Sight, 2D Fresnel Zone, and 3D Fresnel Zone Analysis', *ISPRS International Journal of Geo-Information*, vol. 7, no. 8, Art. no. 8, Aug. 2018, doi: 10.3390/ijgi7080322.

[31] A. Osterman and P. Ritosa, 'Radio Propagation Calculation: A Technique Using 3D Fresnel Zones for Decimeter Radio Waves on Lidar Data', *IEEE Antennas and Propagation Magazine*, vol. 61, no. 6, pp. 31–43, Dec. 2019, doi: 10.1109/MAP.2019.2943312.

[32] D. Townend, S. D. Walker, A. Sharples, and A. Sutton, 'A Unified Line-of-Sight Probability Model for Commercial 5G Mobile Network Deployments', *IEEE Transactions on Antennas and Propagation*, pp. 1–1, 2021, doi: 10.1109/TAP.2021.3119099.

[33] D. González G., H. Hakula, A. Rasila, and J. Hämäläinen, 'Spatial Mappings for Planning and Optimization of Cellular Networks', *IEEE/ACM Transactions on Networking*, vol. 26, no. 1, pp. 175–188, Feb. 2018, doi: 10.1109/TNET.2017.2768561.

[34] A. Taufique, M. Jaber, A. Imran, Z. Dawy, and E. Yacoub, 'Planning Wireless Cellular Networks of Future: Outlook, Challenges and Opportunities', *IEEE Access*, vol. 5, pp. 4821–4845, 2017, doi: 10.1109/ACCESS.2017.2680318.

[35] M. Mozaffari, A. T. Z. Kasgari, W. Saad, M. Bennis, and M. Debbah, 'Beyond 5G With UAVs: Foundations of a 3D Wireless Cellular Network', *IEEE Transactions on Wireless Communications*, vol. 18, no. 1, pp. 357–372, Jan. 2019, doi: 10.1109/TWC.2018.2879940.

[36] S. Zhang and R. Zhang, 'Radio Map-Based 3D Path Planning for Cellular-Connected UAV', *IEEE Transactions on Wireless Communications*, vol. 20, no. 3, pp. 1975–1989, Mar. 2021, doi: 10.1109/TWC.2020.3037916.

[37] J. Wu, P. Yu, L. Feng, F. Zhou, W. Li, and X. Qiu, '3D Aerial Base Station Position Planning based on Deep Q-Network for Capacity Enhancement', in *2019 IFIP/IEEE Symposium on Integrated Network and Service Management (IM)*, Apr. 2019, pp. 482–487.

[38] S. D. Bast, E. Vinogradov, and S. Pollin, 'Cellular Coverage-Aware Path Planning for UAVs', in *2019 IEEE 20th International Workshop on Signal Processing Advances in Wireless Communications (SPAWC)*, Jul. 2019, pp. 1–5. doi: 10.1109/SPAWC.2019.8815469.

[39] U. Masood, H. Farooq, and A. Imran, 'A Machine Learning Based 3D Propagation Model for Intelligent Future Cellular Networks', in *2019 IEEE Global Communications Conference (GLOBECOM)*, Dec. 2019, pp. 1–6. doi: 10.1109/GLOBECOM38437.2019.9014187.

[40] C.-H. Liu, D.-C. Liang, M. A. Syed, and R.-H. Gau, 'A 3D Tractable Model for UAV-Enabled Cellular Networks With Multiple Antennas', *IEEE Transactions on Wireless Communications*, pp. 1–1, 2021, doi: 10.1109/TWC.2021.3051415.

[41] M. Jaber, M. A. Imran, R. Tafazolli, and A. Tukmanov, '5G Backhaul Challenges and Emerging Research Directions: A Survey', *IEEE Access*, vol. 4, pp. 1743–1766, 2016, doi: 10.1109/ACCESS.2016.2556011.

[42] Y. Ren *et al.*, 'Line-of-Sight Millimeter-Wave Communications Using Orbital Angular Momentum Multiplexing Combined With Conventional Spatial Multiplexing', *IEEE Transactions on Wireless Communications*, vol. 16, no. 5, pp. 3151–3161, May 2017, doi: 10.1109/TWC.2017.2675885.

[43] M. Jasim and N. Ghani, 'Sidelobe Exploitation for Beam Discovery in Line of Sight Millimeter Wave Systems', *IEEE Wireless Communications Letters*, vol. 7, no. 2, pp. 234–237, Apr. 2018, doi: 10.1109/LWC.2017.2768528.

[44] L. Zhu, S. Wang, and J. Zhu, 'Adaptive Beamforming Design for Millimeter-Wave Line-of-Sight MIMO Channel', *IEEE Communications Letters*, vol. 23, no. 11, pp. 2095–2098, Nov. 2019, doi: 10.1109/LCOMM.2019.2936379.

[45] B. Malila, O. Falowo, and N. Ventura, 'Intelligent NLOS Backhaul for 5G Small Cells', *IEEE Communications Letters*, vol. 22, no. 1, pp. 189–192, Jan. 2018, doi: 10.1109/LCOMM.2017.2754264.

[46] A. Alorainy and M. J. Hossain, 'Cross-Layer Performance of Channel Scheduling Mechanisms in Small-Cell Networks With Non-Line-of-Sight Wireless Backhaul Links', *IEEE Transactions on Wireless Communications*, vol. 14, no. 9, pp. 4907–4922, Sep. 2015, doi: 10.1109/TWC.2015.2429578.

[47] M. N. Islam, A. Sampath, A. Maharshi, O. Koymen, and N. B. Mandayam, 'Wireless backhaul node placement for small cell networks', in *2014 48th Annual Conference on Information Sciences and Systems (CISS)*, Mar. 2014, pp. 1–6. doi: 10.1109/CISS.2014.6814156.

[48] Y. Zhao, Z. Li, B. Hao, and J. Shi, 'Sensor Selection for TDOA-Based Localization in Wireless Sensor Networks With Non-Line-of-Sight Condition', *IEEE Transactions on Vehicular Technology*, vol. 68, no. 10, pp. 9935–9950, Oct. 2019, doi: 10.1109/TVT.2019.2936110.

[49] X. Sun *et al.*, 'Non-line-of-sight methodology for high-speed wireless optical communication in highly turbid water', *Optics Communications*, vol. 461, p. 125264, Apr. 2020, doi: 10.1016/j.optcom.2020.125264.

[50] Z. Cao, X. Zhang, G. Osnabrugge, J. Li, I. M. Vellekoop, and A. M. J. Koonen, 'Reconfigurable beam system for non-line-of-sight free-space optical communication', *Light: Science & Applications*, vol. 8, no. 1, Art. no. 1, Jul. 2019, doi: 10.1038/s41377-019-0177-3.

[51] J. Hua, Y. Yin, A. Wang, Y. Zhang, and W. Lu, 'Geometry-based non-line-of-sight error mitigation and localization in wireless communications', *Sci. China Inf. Sci.*, vol. 62, no. 10, p. 202301, Aug. 2019, doi: 10.1007/s11432-019-9909-5.

[52] S. Zheng *et al.*, 'Non-Line-of-Sight Channel Performance of Plane Spiral Orbital Angular Momentum MIMO Systems', *IEEE Access*, vol. 5, pp. 25377–25384, 2017, doi: 10.1109/ACCESS.2017.2766078.

[53] B. Olsson, C. Larsson, and J. Hansryd, 'Angular Resolved Site Characterization of Non-Line-of-Sight Wireless Links', in *2017 IEEE Wireless Communications and Networking Conference (WCNC)*, Mar. 2017, pp. 1–5. doi: 10.1109/WCNC.2017.7925956.

[54] H. Dahrouj, A. Douik, F. Rayal, T. Y. Al-Naffouri, and M.-S. Alouini, 'Cost-effective hybrid RF/FSO backhaul solution for next generation wireless systems', *IEEE Wireless Communications*, vol. 22, no. 5, pp. 98–104, Oct. 2015, doi: 10.1109/MWC.2015.7306543.

[55] M. Coldrey, J. Berg, L. Manholm, C. Larsson, and J. Hansryd, 'Non-line-of-sight small cell backhauling using microwave technology', *IEEE Communications Magazine*, vol. 51, no. 9, pp. 78–84, Sep. 2013, doi: 10.1109/MCOM.2013.6588654.

[56] A. Chehri and H. T. Mouftah, 'New MMSE Downlink Channel Estimation for Sub-6 GHz Non-Line-of-Sight Backhaul', in *2018 IEEE Globecom Workshops (GC Wkshps)*, Dec. 2018, pp. 1–7. doi: 10.1109/GLOCOMW.2018.8644436.

[57] S. Rajagopal, S. Abu-Surra, and M. Malmirchegini, 'Channel Feasibility for Outdoor Non-Line-of-Sight mmWave Mobile Communication', in *2012 IEEE Vehicular Technology Conference (VTC Fall)*, Sep. 2012, pp. 1–6. doi: 10.1109/VTCFall.2012.6398884.

[58] H. B. H. Dutty and M. M. Mowla, 'Channel Modeling at Unlicensed Millimeter Wave V Band for 5G Backhaul Networks', in *2019 5th International Conference on Advances in Electrical Engineering (ICAEE)*, Sep. 2019, pp. 769–773. doi: 10.1109/ICAEE48663.2019.8975439.

[59] C. Zhang, H. Wu, H. Lu, and J. Liu, 'Throughput Analysis in Cache-enabled Millimeter Wave HetNets with Access and Backhaul Integration', in *2020 IEEE Wireless Communications and Networking Conference (WCNC)*, May 2020, pp. 1–6. doi: 10.1109/WCNC45663.2020.9120585.

[60] S. Pérez-Peña *et al.*, 'Development of Measurement and Modeling Procedures of Diffractive near-LOS Wireless Links', in *2020 XXXIIIrd General Assembly and Scientific Symposium of the International Union of Radio Science*, Aug. 2020, pp. 1–4. doi: 10.23919/URSIGASS49373.2020.9232262.

[61] G. Yue, D. Yu, K. Qiu, K. Guan, L. Yang, and Q. Lv, 'Measurements and Ray Tracing Simulations for Non-Line-of-Sight Millimeter-Wave Channels in a Confined Corridor Environment', *IEEE Access*, vol. 7, pp. 85066–85081, 2019, doi: 10.1109/ACCESS.2019.2924510.

[62] Cloud-RF, 'Online RF planning software', *Cloud-RF™*, 2021. https://cloudrf.com/ (accessed Nov. 14, 2021).

[63] Forsk, 'Radio Planning and Optimisation Software', 2021. https://www.forsk.com/ (accessed Nov. 14, 2021).

[64] Splat, 'A Terrestrial RF Path Analysis Application For Linux/Unix', 2021. https://www.qsl.net/kd2bd/splat.html (accessed Nov. 14, 2021).

[65] Pathloss, 'Microwave Radio Link Design and Planning Software', 2021. https://www.pathloss.com/#!main (accessed Nov. 14, 2021).

[66] MathWorks, 'MATLAB Communications Toolbox', 2021. https://www.mathworks.com/products/communications.html (accessed Nov. 14, 2021).

[67] EDX, 'Best Radio / Wireless Planning Software', *EDX Wireless*, 2021. https://edx.com/ (accessed Nov. 14, 2021).

[68] Remcom, 'Electromagnetic Simulation Software & EM Modeling', *Remcom*, 2021. https://www.remcom.com (accessed Nov. 14, 2021).

[69] Facebook, 'Announcing tools to help partners improve connectivity', *Facebook Engineering*, Aug. 10, 2018. https://engineering.fb.com/2018/08/10/connectivity/announcing-tools-to-help-partners-improve-connectivity/ (accessed Nov. 14, 2021).

[70] E. J. Oughton, *edwardoughton/e3nb*. Fairfax, VA: George Mason University, 2021. Accessed: Apr. 22, 2021. [Online]. Available: https://github.com/edwardoughton/e3nb

[71] International Telecommunication Union, 'Propagation data and prediction methods required for the design of terrestrial line-of-sight systems. Recommendation itu-r p.530-10.', International Telecommunication Union, Geneva, Switzerland, 2001.

[72] International Telecommunication Union, 'ITU-R P.526-15 : Propagation by diffraction', International Telecommunication Union, Geneva, Switzerland, 2019. Accessed: May 14, 2021. [Online]. Available: https://www.itu.int/rec/R-REC-P.526-15-201910-I/en

[73] H. Lehpamer, *Microwave transmission networks: planning, design, and deployment*. McGraw-Hill Education, 2010.

[74] O. Taghizadeh, P. Sirvi, S. Narasimha, J. A. L. Calvo, and R. Mathar, 'Environment-Aware Minimum-Cost Wireless Backhaul Network Planning

With Full-Duplex Links', *IEEE Systems Journal*, vol. 13, no. 3, pp. 2582–2593, Sep. 2019, doi: 10.1109/JSYST.2019.2893537.

[75] A. Mahmood *et al.*, 'Capacity and Frequency Optimization of Wireless Backhaul Network Using Traffic Forecasting', *IEEE Access*, vol. 8, pp. 23264–23276, 2020, doi: 10.1109/ACCESS.2020.2970224.

[76] G. Zhang, T. Q. S. Quek, M. Kountouris, A. Huang, and H. Shan, 'Fundamentals of Heterogeneous Backhaul Design—Analysis and Optimization', *IEEE Transactions on Communications*, vol. 64, no. 2, pp. 876–889, Feb. 2016, doi: 10.1109/TCOMM.2016.2515596.

[77] W. Ding, Y. Niu, H. Wu, Y. Li, and Z. Zhong, 'QoS-Aware Full-Duplex Concurrent Scheduling for Millimeter Wave Wireless Backhaul Networks', *IEEE Access*, vol. 6, pp. 25313–25322, 2018, doi: 10.1109/ACCESS.2018.2828852.

[78] M. Khaturia, K. Appaiah, and A. Karandikar, 'On Efficient Wireless Backhaul Planning for the "Frugal 5G" Network', in *2019 IEEE Wireless Communications and Networking Conference Workshop (WCNCW)*, Apr. 2019, pp. 1–6. doi: 10.1109/WCNCW.2019.8902828.

[79] Y. Huang, M. Cui, G. Zhang, and W. Chen, 'Bandwidth, Power and Trajectory Optimization for UAV Base Station Networks With Backhaul and User QoS Constraints', *IEEE Access*, vol. 8, pp. 67625–67634, 2020, doi: 10.1109/ACCESS.2020.2986075.

[80] G. Destino *et al.*, 'System analysis and design of mmW mobile backhaul transceiver at 28 GHz', in *2017 European Conference on Networks and Communications (EuCNC)*, Jun. 2017, pp. 1–5. doi: 10.1109/EuCNC.2017.7980768.

[81] GSMA, 'The 5G guide: A reference for operators', GSMA, London, 2019.

[82] T. Manning, *Microwave radio transmission design guide*. Artech house, 2009.

[83] L. Chiaraviglio, C. Di Paolo, and N. Blefari Melazzi, '5G Network Planning under Service and EMF Constraints: Formulation and Solutions', *IEEE Transactions on Mobile Computing*, pp. 1–1, 2021, doi: 10.1109/TMC.2021.3054482.

[84] I. Mesogiti *et al.*, 'Macroscopic and microscopic techno-economic analyses highlighting aspects of 5G transport network deployments', *Photon Netw Commun*, vol. 40, no. 3, pp. 256–268, Dec. 2020, doi: 10.1007/s11107-020-00912-w.

[85] E. J. Oughton and A. Jha, 'Supportive 5G Infrastructure Policies are Essential for Universal 6G: Assessment Using an Open-Source Techno-Economic Simulation Model Utilizing Remote Sensing', *IEEE Access*, vol. 9, pp. 101924–101945, 2021, doi: 10.1109/ACCESS.2021.3097627.

[86] Ofcom, 'Mobile call termination market review 2018-21', *Ofcom*, Mar. 28, 2018.

https://www.ofcom.org.uk/consultations-and-statements/category-1-mobile-call-termination-market-review (accessed Sep. 09, 2019).

[87] H. Holma and A. Toskala, Eds., *LTE-Advanced: 3GPP Solution for IMT-Advanced*. Chichester, UK: John Wiley & Sons, Ltd, 2012. Accessed: Oct. 25, 2016. [Online]. Available: http://doi.wiley.com/10.1002/9781118399439

[88] SEAMCAT, *Spectrum Engineering Advanced Monte Carlo Analysis Tool (SEAMCAT) Handbook*. SEAMCAT, 2010. Accessed: Oct. 31, 2016. [Online]. Available: http://www.seamcat.org/

[89] Dragonwave, 'Horizon Compact. Wireless Ethernet Release 1.01.01. Product Manual - Volume 1. Version 1.4.', Dragonwave, Ottawa, Canada, Product Manual 1.01.01, 2008. [Online]. Available: https://fccid.io/QB8HC-24UL/User-Manual/User-Manual-908839.pdf?fbclid=IwAR0KgS8qlBekkr7yZYE6skFR89oPMDPY9dkUqN_Sn1uEGRS8yeb47kZlrBo

[90] A. Beltramo, E. P. Ramos, C. Taliotis, M. Howells, and W. Usher, 'The Global Least-cost user-friendly CLEWs Open-Source Exploratory model', *Environmental Modelling & Software*, vol. 143, p. 105091, Sep. 2021, doi: 10.1016/j.envsoft.2021.105091.

[91] D. Mentis *et al.*, 'Lighting the World: the first application of an open source, spatial electrification tool (OnSSET) on Sub-Saharan Africa', *Environ. Res. Lett.*, vol. 12, no. 8, p. 085003, Jul. 2017, doi: 10.1088/1748-9326/aa7b29.

[92] B. Khavari, A. Korkovelos, A. Sahlberg, M. Howells, and F. Fuso Nerini, 'Population cluster data to assess the urban-rural split and electrification in Sub-Saharan Africa', *Scientific Data*, vol. 8, no. 1, Art. no. 1, Apr. 2021, doi: 10.1038/s41597-021-00897-9.

[93] E. Oughton, 'Policy options for digital infrastructure strategies: A simulation model for broadband universal service in Africa', *arXiv:2102.03561 [cs, econ, q-fin]*, Feb. 2021, Accessed: Feb. 09, 2021. [Online]. Available: http://arxiv.org/abs/2102.03561

[94] A. Getis, 'Spatial Autocorrelation', in *Handbook of Applied Spatial Analysis: Software Tools, Methods and Applications*, M. M. Fischer and A. Getis, Eds. Berlin, Heidelberg: Springer, 2010, pp. 255–278. doi: 10.1007/978-3-642-03647-7_14.

[95] A. Getis, 'Reflections on spatial autocorrelation', *Regional Science and Urban Economics*, vol. 37, no. 4, pp. 491–496, Jul. 2007, doi: 10.1016/j.regsciurbeco.2007.04.005.

[96] B. Khavari, A. Sahlberg, W. Usher, A. Korkovelos, and F. Fuso Nerini, 'The effects of population aggregation in geospatial electrification planning', *Energy Strategy Reviews*, vol. 38, p. 100752, Nov. 2021, doi: 10.1016/j.esr.2021.100752.

[97] M. Ceci, R. Corizzo, D. Malerba, and A. Rashkovska, 'Spatial autocorrelation and entropy for renewable energy forecasting', *Data Min Knowl Disc*, vol. 33, no. 3, pp. 698–729, May 2019, doi: 10.1007/s10618-018-0605-7.

[98] S. Thomas, I. K. Gayathri, and A. Raj, 'Joint design of Dijkstra's shortest path routing and sleep-wake scheduling in wireless sensor networks', in *2017 International Conference on Energy, Communication, Data Analytics and Soft Computing (ICECDS)*, Aug. 2017, pp. 981–986. doi: 10.1109/ICECDS.2017.8389583.

[99] H. Juzoji, I. Nakajima, and T. Kitano, 'A Development of Network Topology of Wireless Packet Communications for Disaster Situation with Genetic Algorithms or with Dijkstra's's', in *2011 IEEE International Conference on Communications (ICC)*, Jun. 2011, pp. 1–5. doi: 10.1109/icc.2011.5962439.

[100] N.-H. Bao, D.-Y. Luo, and J.-B. Chen, 'Reliability threshold based service bandwidth recovery scheme for post-disaster telecom networks', *Optical Fiber Technology*, vol. 45, pp. 81–88, Nov. 2018, doi: 10.1016/j.yofte.2018.06.008.

[101] E. J. Oughton, T. Russell, J. Johnson, C. Yardim, and J. Kusuma, 'itmlogic: The Irregular Terrain Model by Longley and Rice', *Journal of Open Source Software*, vol. 5, no. 51, p. 2266, Jul. 2020, doi: 10.21105/joss.02266.

[102] G. A. Hufford, A. G. Longley, and W. A. Kissick, *A guide to the use of the ITS irregular terrain model in the area prediction mode*. US Department of Commerce, National Telecommunications and Information …, 1982.

[103] G. A. Hufford, 'The ITS irregular terrain model', *Institute for Telecommunication Sciences, National Telecommunications and Information Administration, Boulder, CO, USA*, 1995.

[104] European Telecommunications Standards Institute, 'ETSI TR 138 901 V15.0.0 (2018-07). 5G; Study on channel model for frequencies from 0.5 to 100 GHz (3GPP TR 38.901 version 15.0.0 Release 15)', ETSI, Sophia Antipolis, France, 2018.

[105] International Telecommunication Union, 'Characteristics of precipitation for propagation modelling. Recommendation itu-r pn.837-1.', International Telecommunication Union, Geneva, Switzerland, 1992. [Online]. Available: https://www.itu.int/dms_pubrec/itu-r/rec/p/R-REC-P.837-1-199408-S!!PDF-E.pdf

[106] Y. Zhang, C. R. Anderson, N. Michelusi, D. J. Love, K. R. Baker, and J. V. Krogmeier, 'Propagation Modeling Through Foliage in a Coniferous Forest at 28 GHz', *IEEE Wireless Communications Letters*, vol. 8, no. 3, pp. 901–904, Jun. 2019, doi: 10.1109/LWC.2019.2899299.

[107] R. Anzum, M. H. Habaebi, M. R. Islam, and G. P. N. Hakim, 'Modeling and Quantifying Palm Trees Foliage Loss using LoRa Radio Links for Smart Agriculture Applications', in *2021 IEEE 7th*

International Conference on Smart Instrumentation, Measurement and Applications (ICSIMA)*, Aug. 2021, pp. 105–110. doi: 10.1109/ICSIMA50015.2021.9526311.

[108] SWG, 'EtherHaul-2500FX ODU', *SWG Inc.*, 2021. https://swginc.com/product/etherhaul-2500fx-odu-with-aes-hw-license-with-ant-port-tx-low-power-poedc-1g-upgradable-to-2ge-ports2xcopper-2xfiber-high-power/ (accessed Nov. 10, 2021).

[109] SWG, '0.6m (2') E-Band Microwave Antenna Series', *SWG Inc.*, 2021. https://swginc.com/product/rf-engineering-energy-resource-rfe-80-ghz-71-0-86-0ghz-0-6m-2-e-band-microwave-antenna-series-ultra-high-performance-single-polarized-direct-connect-remec-interface/ (accessed Nov. 10, 2021).

[110] SWG, '0.9m (3') Universal Microwave Antenna Series', *SWG Inc.*, 2021. https://swginc.com/product/rf-engineering-energy-resource-rfe-11-ghz-10-125-11-70ghz-0-9m-3-universal-microwave-antenna-series-ultra-high-performance-dual-polarized-ceragon-ip-20c-interface/ (accessed Nov. 10, 2021).

[111] SWG, '1.2m (4') Universal Microwave Antenna', *SWG Inc.*, 2021. https://swginc.com/product/rf-engineering-energy-resource-rfe-13-ghz-12-75-13-25ghz-1-2m-4-universal-microwave-antenna-series-ultra-high-performance-single-polarized-12-75-13-25ghz-waveguide-interface/ (accessed Nov. 10, 2021).

[112] SWG, '1.8m (6') Universal Microwave Antenna Series', *SWG Inc.*, 2021. https://swginc.com/product/rf-engineering-energy-resource-rfe-18-ghz-17-7-19-7ghz-1-8m-6-universal-microwave-antenna-series-ultra-high-performance-single-polarized-waveguide-interface/ (accessed Nov. 10, 2021).

[113] O. Tupayachi, 'Challenges of rural broadband coverage in Peru', presented at the Facebook Connectivity Workshop 2019, Menlo Park, CA, USA, 2019.

[114] M. Paolini, 'Crucial economics for mobile data backhaul. An analysis of the total cost of ownership of point-to-point, point-to-multipoint, and fibre options', Senza Fili, Seattle, WA, USA, 2011. [Online]. Available: https://www.alliancecorporation.ca/images/documents/broadband-documents/Cambridge_Broadband/Cambridge-Broadband-Whitepaper-Crucial-economics-for-mobile-data-backhaul.pdf

[115] Alibaba, 'Mobile Solar Power Trailer System', *Alibaba.com*, 2021. https://www.alibaba.com/product-detail/mobile-solar-power-trailer-systems_60478439962.html?spm=a2700.7724857.no

rmal_offer.d_image.6ed3124amgv78Z (accessed Nov. 10, 2021).

[116] M. Mahloo, P. Monti, J. Chen, and L. Wosinska, 'Cost modeling of backhaul for mobile networks', in *2014 IEEE International Conference on Communications Workshops (ICC)*, 2014, pp. 397–402. Accessed: Oct. 25, 2016. [Online]. Available: http://ieeexplore.ieee.org/xpls/abs_all.jsp?arnumber=6881230

[117] GSMA, 'Wireless backhaul evolution. Delivering next-generation connectivity.', GSMA, London, 2021. [Online]. Available: https://www.gsma.com/spectrum/wp-content/uploads/2021/02/wireless-backhaul-spectrum.pdf

[118] GADM, 'Global Administrative Areas Database (Version 3.6)', 2019. https://gadm.org/ (accessed Jul. 11, 2019).

[119] WorldPop, 'WorldPop :: Population', 2019. https://www.worldpop.org/project/categories?id=3 (accessed Jan. 02, 2020).

[120] A. J. Tatem, 'WorldPop, open data for spatial demography', *Sci Data*, vol. 4, no. 1, pp. 1–4, Jan. 2017, doi: 10.1038/sdata.2017.4.

[121] United States Geological Survey, 'Global Multi-resolution Terrain Elevation Data (GMTED2010)', United States Geological Survey, Reston, VA., 2010.

[122] DiMiceli, C., Carroll, M., Sohlberg, R., Kim, D., Kelly, M., and Townshend, J., 'MOD44B MODIS/Terra Vegetation Continuous Fields Yearly L3 Global 250m SIN Grid V006'. NASA EOSDIS Land Processes DAAC, 2015. doi: 10.5067/MODIS/MOD44B.006.

[123] S. P. Healey *et al.*, 'CMS: GLAS LiDAR-derived Global Estimates of Forest Canopy Height, 2004-2008', *ORNL DAAC*, 2016, doi: 10.3334/ORNLDAAC/1271.

[124] X. Lu, M. Salehi, M. Haenggi, E. Hossain, and H. Jiang, 'Stochastic Geometry Analysis of Spatial-Temporal Performance in Wireless Networks: A Tutorial', *IEEE Communications Surveys Tutorials*, pp. 1–1, 2021, doi: 10.1109/COMST.2021.3104581.

[125] A. Guo and M. Haenggi, 'Spatial Stochastic Models and Metrics for the Structure of Base Stations in Cellular Networks', *IEEE Transactions on Wireless Communications*, vol. 12, no. 11, pp. 5800–5812, Nov. 2013, doi: 10.1109/TWC.2013.100113.130220.

[126] H. Chen, L. Liu, H. S. Dhillon, and Y. Yi, 'QoS-Aware D2D Cellular Networks With Spatial Spectrum Sensing: A Stochastic Geometry View', *IEEE Transactions on Communications*, vol. 67, no. 5, pp. 3651–3664, May 2019, doi: 10.1109/TCOMM.2018.2889246.

[127] F. Ademaj, S. Schwarz, T. Berisha, and M. Rupp, 'A Spatial Consistency Model for Geometry-Based Stochastic Channels', *IEEE Access*, vol. 7, pp.

183414–183427, 2019, doi:
10.1109/ACCESS.2019.2958154.

[128] D. Dupleich, H. Abbas Mir, C. Schneider, G. Del
Galdo, and R. Thomä, 'On the Modelling of the
NLOS First Multi-path Component in Stochastic
Spatial Channel Models', in *2021 15th European
Conference on Antennas and Propagation (EuCAP)*,
Mar. 2021, pp. 1–5. doi:
10.23919/EuCAP51087.2021.9411142.

[129] S. Dang, O. Amin, B. Shihada, and M.-S. Alouini,
'What should 6G be?', *Nat Electron*, vol. 3, no. 1, pp.
20–29, Jan. 2020, doi: 10.1038/s41928-019-0355-6.